# Insight into magnetocaloric properties of Mn$_2$Nb molecular magnet by relaxation calorimetry: A comprehensive case study

Robert Pełka[1,*], Yuji Miyazaki[2], Yasuhiro Nakazawa[3], Dawid Pinkowicz[4]

[1] The H. Niewodniczański Institute of Nuclear Physics Polish Academy of Sciences, Radzikowskiego 152, 31-342 Kraków, Poland

[2] Research Center for Thermal and Entropic Science, Graduate School of Science, Osaka University, Toyonaka, Osaka 560-0043, Japan

[3] Department of Chemistry, Graduate School of Science, Osaka University, Machikaneyama 1-1, Toyonaka, Osaka 560-0043, Japan

[4] Faculty of Chemistry, Jagiellonian University, Gronostajowa 2, 30-387 Kraków, Poland

* Corresponding author, e-mail: robert.pelka@ifj.edu.pl

**Abstract**

Magnetocaloric effect in [Nb$^{IV}${($\mu$-CN)$_4$Mn$^{II}$(H$_2$O)$_2$]}$_2$·4H$_2$O]$_n$ molecular magnet is reported. The compound crystallizes in the tetragonal I4/m space group. It exhibits a phase transition to a long-range ferrimagnetically ordered state at $T_c$ = 47.0(2) K. In order to calculate magnetocaloric properties relaxation calorimetry measurements are performed and a self-consistent scheme based on the magnetic entropy counting for the baseline determination is developed. The magnetic entropy change $\Delta S_M$ as well as the adiabatic temperature change $\Delta T_{ad}$ due to the applied field changes $\mu_0\Delta H$ = 0.1, 0.2, 0.5, 1, 2, 3, 4, 5, 7, and 9 T as functions of temperature are evaluated. The maximum value of $|\Delta S_M|$ for $\mu_0\Delta H$ = 5 T is 5.03 J K$^{-1}$ mol$^{-1}$ (9.07 J K$^{-1}$ kg$^{-1}$) at 49.5 K. The corresponding maximum value of $\Delta T_{ad}$ = 1.7 K is attained at 49.0 K. The molecular field model is used to simulate the temperature and field dependence of the magnetic entropy change. The exchange coupling constant between the Mn$^{II}$ and Nb$^{IV}$ ions is estimated to be equal to -10.26 K. At the lowest temperatures and for the lowest applied field change values the inverse magnetocaloric effect is revealed, which seems to be characteristic for systems with antiferromagnetic coupling. The temperature dependence of exponent $n$ quantifying the field dependence of $\Delta S_M$ is calculated on the basis of the experimental results and within the mean-field model. Its predicting power for the universality class of the critical behavior is discussed. Finally, the studied compound is employed as the working substance in the two most natural refrigeration cycles, i.e. the Brayton cycle and the Ericsson cycle, to assess its cooling effectiveness. A cascade system is suggested for the most efficient cooling performance.

Keywords: heat capacity, magnetocaloric effect, entropy, molecular-field approximation, refrigeration cycles

1. Introduction

Molecular magnetism has been enjoying a dynamical development for several decades now [1-5]. From its very beginning it was distinguished by a close cooperation between chemists and physicists who joined their competences to create and thoroughly characterize new molecular-based compounds representing a promising alternative to conventional magnets [6]. Concerted efforts of the researchers led to the groundbreaking discoveries of single molecule magnets (SSM) [7-9] and multifunctional materials such as stimuli-responsive magnets [10-20], porous magnets [21-24], molecular conducting magnets [25-27], chiral magnets [28-35], and magnetic superconductors [36-38]. Recent advances in this field include new research areas such as molecular spintronics, quantum computing, and the construction of novel metal-organic frameworks (MOFs) and two-dimensional (2D) molecular materials [39-57]. The subfield of multifunctional molecular magnets still being the focus of the scientists' attention [58-62], it is also joined by the present report concerning the potential application of a novel magnetic coordination polymer as a working substance in cryogenic cooling devices. The phenomenon underlying this technology is the magnetocaloric effect (MCE) consisting in a magnetic substance heating/cooling due to the switching on/off the external magnetic field, in which case one says about the direct or normal MCE [63-67]. The magnetocaloric refrigeration owes its enduring interest to the fact that it is an environmentally friendly method with performance competitive with conventional cooling methods based on gas compression and expansion. Usually the performance of a magnetic material is described by determining two interrelated quantities, i.e., the isothermal magnetic entropy change $\Delta S_M$ and the adiabatic temperature change $\Delta T_{ad}$ associated with external field change $H_i \to H_f$, where the lower indices stand for *initial* and *final*, respectively. While the former quantity can be readily calculated on the basis of the temperature and field evolution of the magnetization (indirect magnetometric method) by using the integral version of the thermodynamic Maxwell relation, i.e.,

$$\Delta S_M(T, H_i \to H_f) = \mu_0 \int_{H_i}^{H_f} \frac{\partial M(T,H)}{\partial T} dH, \qquad (1)$$

$\Delta T_{ad}$ requires a measurement of the field and temperature dependence of the heat capacity (indirect calorimetric method). The adiabatic temperature change is related to the isothermal entropy change, i.e.,

$$\delta T_{ad}(T, \delta H) = -\frac{T}{C_p(T,H)} \delta S_M(T, \delta H), \qquad (2)$$

where $\delta S_M$ and $\delta T_{ad}$ denote infinitesimal changes associated with the infinitesimal change of the field $\delta H = H - H_i$, and $C_p$ is the heat capacity at constant pressure. Out of the two methods the calorimetric one is more universal as it allows for accurate determination of both the magnetocaloric quantities. For this very reason the latter method is used in the present study. For the sake of completeness, let us mention that there are accounts in the literature of direct measurement of the adiabatic temperature change [68-73]. However, such a measurement requires a specialized, commercially unavailable device adapted to detect temperature changes under conditions of thermal insulation.

The quest for novel magnetocaloric materials among molecular magnets is an ongoing process [74-77]. Their investigation goes mainly along experimental lines [78-96], but there are also theoretical accounts [97-103]. The studied compound belongs to an important subclass of molecular magnets in which the magnetic ions are interconnected by inorganic cyanido bridges –CN– offering a relatively strong exchange coupling. We have been investigating MCE of these materials for over a decade now and the results obtained may serve as the natural reference point for the current research work [104-119]. The corresponding compounds accompanied with the most important MCE characteristics are collected in Table 1. Most of them are two-dimensional (2D) or three-dimensional (3D) coordination polymers displaying the second order transition to a magnetically ordered phase with transition temperatures ranging from 4.8 to 95.2 K. The corresponding peak values of the isothermal entropy change $\Delta S_\text{M}^\text{peak}$ due to switching off the field of the magnitude of 5 T span the interval from 1.41 up to 7.04 J K$^{-1}$ mol$^{-1}$. Besides, there are two instances of fifteen-center spin clusters (cf. No. 10 and 12 in Table 1) with intermolecular interactions too weak to lead to magnetic ordering. The maximal isothermal entropy changes for $\mu_0 \Delta H = -5$ T are in this case greater reflecting high-resultant-spin paramagnetic nature of the systems. Not all the compounds were measured by a calorimetric method allowing one to quantify the adiabatic change of temperature due to switching on the magnetic field of 5 T (cf. column 5 in Table 1). The reported values of $\Delta T_\text{ad}^\text{peak}$ range from 0.4 to 4.6 K and were found to depend significantly on the molecular complexity of the material, which is expressed by the magnitude of the heat capacity.

**Table 1:** MCE parameters for a selection of cyanido-bridged molecular magnets.

| No. | Compound | $T_c$ [K] | $\Delta S_\text{M}^\text{peak}$ [J K$^{-1}$mol$^{-1}$] ($\mu_0 \Delta H = -5$ T) | $\Delta T_\text{ad}^\text{peak}$ [K] ($\mu_0 \Delta H = 5$ T) | Ref. |
|---|---|---|---|---|---|
| 1 | [{[Mn$^\text{II}$(pydz)(H$_2$O)$_2$][Mn$^\text{II}$(H$_2$O)$_2$][Nb$^\text{IV}$(CN)$_8$]}·2H$_2$O]$_n$ | 42.1 | 5.36 | 1.5 | [104] |
| 2 | {[Mn$^\text{II}$(pydz)(H$_2$O)][Mn$^\text{II}$(H$_2$O)][Nb$^\text{IV}$(CN)$_8$]}$_n$ | 65.9 | 3.33 | – | [104] |
| 3 | {[Mn$^\text{II}$$_2$(pydz)][Nb$^\text{IV}$(CN)$_8$]}$_n$ | 95.2 | 3.38 | – | [104] |
| 4 | {[Ni$^\text{II}$(pyrazole)$_4$]$_2$[Nb$^\text{IV}$(CN)$_8$]·4H$_2$O | 13.4 | 6.1 | 2.0 | [105] |
| 5 | {[Mn$^\text{II}$(pyrazole)$_4$]$_2$[Nb$^\text{IV}$(CN)$_8$]·4H$_2$O | 22.8 | 6.7 | 1.42 | [105] [111] |
| 6 | {(tetrenH$_5$)$_{0.8}$Cu$^\text{II}$$_4$[W(CN)$_8$]$_4$·7.2H$_2$O}$_n$ | 34 | 3.1 | 0.4 | [107] [112] |
| 7 | {[Fe$^\text{II}$(pyrazole)$_4$]$_2$[Nb$^\text{IV}$(CN)$_8$]·4H$_2$O | 8.3 | 4.9 | 2.0 | [108] [117] |
| 8 | {Mn$^\text{II}$$_2$(imH)$_2$(H$_2$O)$_4$[Nb$^\text{IV}$(CN)$_8$]·4H$_2$O}$_n$ | 24.1 | 6.70 | 2.02 | [109] [119] |
| 9 | {Mn$^\text{II}$$_2$(imH)$_2$[Nb$^\text{IV}$(CN)$_8$]}$_n$ | 60 | 4.02 | – | [109] |
| 10 | {Ni[Ni(4,4'dtbpy)(H$_2$O)]$_8$[W(CN)$_8$]$_6$}·17H$_2$O | – | 18.38 | 4.6 | [110] |
| 11 | {[Mn$^\text{II}$(R-mpm)$_2$]$_2$[Nb$^\text{IV}$(CN)$_8$]·4H$_2$O}$_n$ | 24 | 5.0 | – | [113] |
| 12 | {Mn$^\text{II}$$_9$(4,4'-dpds)$_4$(MeOH)$_{16}$[W$^\text{V}$(CN)$_8$]$_6$}·12MeOH | – | 31.19 | – | [114] |
| 13 | {Fe$^\text{II}$(H$_2$O)$_2$]$_2$[Nb$^\text{IV}$(CN)$_8$]·4H$_2$O}$_n$ | 43 | 4.82 | – | [115] |
| 14 | [Nb$^\text{IV}${($\mu$-CN)$_4$Mn$^\text{II}$(H$_2$O)$_2$}$_2$·4H$_2$O]$_n$ | 50 | 5.07 | – | [115] |
| 15 | {[Co$^\text{II}$(pyrazole)$_4$]$_2$[Nb$^\text{IV}$(CN)$_8$]·4H$_2$O}$_n$ | 4.87 | 7.04 | 4.16 | [117] |
| 16 | {[Co$^\text{II}$(pyrazole)$_4$][Fe$^\text{II}$(pyrazole)$_4$][Nb$^\text{IV}$(CN)$_8$]·4H$_2$O}$_n$ | 7.1 | 5.26 | 2.47 | [117] |
| 17 | (AdeH){Cu$^\text{II}$[W$^\text{V}$(CN)$_8$]}·2H$_2$O | 33 | 1.41 | – | [118] |

Ligands: pydz = pyridazine (C$_4$H$_4$N$_2$); pyrazole = C$_3$H$_4$N$_2$; tetren = tetraethylenepentamine; imH = imidazole (C$_3$H$_4$N$_2$); 4,4'dtbpy = 4-4-ditertbutyl-2,2'-bipyridine (C$_{20}$H$_{24}$N$_2$O$_4$); mpm = α-methyl-2-pyridinemethanol; 4,4'-dpds = 4,4'-dipyridyl disulfide; Ade = adeninium;

The present paper is devoted to a calorimetric study of MCE in a bimetallic molecular magnet [Nb$^{IV}${(μ-CN)$_4$Mn$^{II}$(H$_2$O)$_2$]}$_2$·4H$_2$O]$_n$. The reported heat capacity measurements enable one to derive both the isothermal entropy change $\Delta S_\mathrm{M}$ and the adiabatic temperature change $\Delta T_\mathrm{ad}$ due to magnetic field changes ranging from 0.1 to 9 T. The structure of the studied compound is unique as it does not contain any organic ligands, the only components being the magnetic manganese and niobium ions and the bridging CN$^-$ groups. Indeed, this unblocked magnetic connectivity has already been tested for magnetocaloric performance by the indirect magnetometric method [115], however the present approach based on the relaxation calorimetry will be demonstrated to provide a more comprehensive insight into its MCE properties. The molar magnetic entropy content of the system is $S_\mathrm{M}^\mathrm{max} = R\ln(2S_\mathrm{Mn}+1)^2(2S_\mathrm{Nb}+1) \approx 35.56\ \mathrm{J\,K^{-1}mol^{-1}}$ with the spin carried by the Mn$^{II}$ ion being equal to $S_\mathrm{Mn}=5/2$ and that of the Nb$^{IV}$ ion amounting to $S_\mathrm{Nb}=1/2$, and $R$ denoting the ideal gas constant. Although this value is quite significant, one cannot expect an MCE of such amplitude due to the fact that the coupling between the Mn$^{II}$ and Nb$^{IV}$ ions was found to be antiferromagnetic [120], making the effective spin per formula unit amount to $S_\mathrm{eff} = 2S_\mathrm{Mn} - S_\mathrm{Nb} = 9/2$. Thus the entropy scale to be observed below the decupling threshold is rather given by $R\ln(2S_\mathrm{eff}+1) \approx 19.14\ \mathrm{J\,K^{-1}mol^{-1}}$, which is reduced roughly by half but still considerable.

The paper is organized as follows. Section 2 is devoted to the crystallographic structure of the studied compound. A short Section 3 provides the crucial experimental details. In Section 4 the heat capacity of the compound as obtained by relaxation calorimetry is discussed and a self-consistent procedure to quantify the lattice contribution to the total heat capacity is proposed. Section 5 is devoted to the evaluation and discussion of the magnetocaloric effect in terms of both the isothermal entropy change and the adiabatic temperature change. The field dependence of the isothermal entropy change is analyzed in detail in Section 6. Section 7 includes the analysis of two thermodynamic cycles with the studied compound employed as the working substance. The paper is wound up by Section 8 listing a number of remarks and conclusions.

2. Structure

The studied compound, [Nb$^{IV}${(μ-CN)$_4$Mn$^{II}$(H$_2$O)$_2$]}$_2$·4H$_2$O]$_n$, crystallizes in the tetragonal space group I4/m with $a = b = 12.080(2)$ Å, $c = 13.375(4)$ Å, and $\alpha = \beta = \gamma = 90°$ [120]. Figure 1 shows its crystal structure seen along the $c$ crystallographic axis. The cyanido bridges can be seen to connect the Mn$^{II}$ ions with the Nb$^{IV}$ ions in a ratio 2:1 with both positions of the Mn$^{II}$ ion being crystallographically equivalent (symmetry related). There is a fourfold axis passing through each Nb$^{IV}$ center and a mirror plane defined by each Mn$^{II}$ center and two coordinated oxygen atoms. Figures 2 and 3 show the coordination spheres of the Nb$^{IV}$ and Mn$^{II}$ ions, respectively. The coordination sphere of the Nb$^{IV}$ ion comprises eight cyanido ligands forming a distorted square antiprism with the carbon atoms pointing towards the central ion. Each Mn$^{II}$ ion coordinates four nitrogen atoms from the cyanido bridges binding in equatorial positions and two oxygen atoms from the water molecules occupying the axial positions. The corresponding coordination sphere has a slightly distorted octahedral geometry.

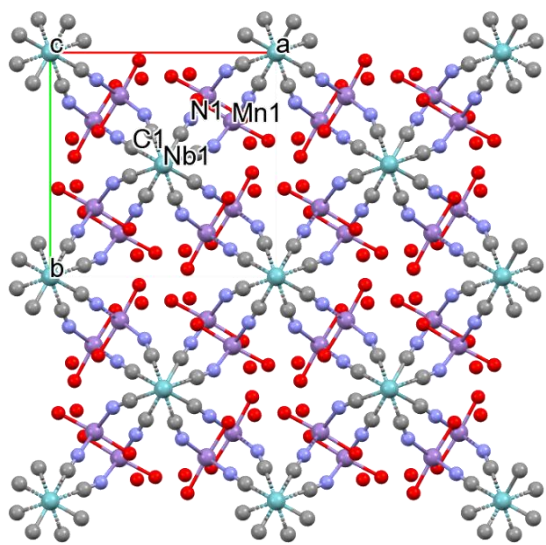

**Fig 1:** The crystal structure of the studied compound as seen along the *c* crystallographic axis. For the sake of clarity the hydrogen atoms have been omitted.

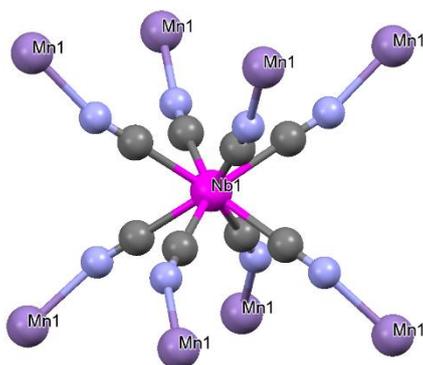

**Fig. 2:** The coordination sphere of the $Nb^{IV}$ ion comprises eight carbon atoms from the cyanido linkages forming an approximate square antiprism.

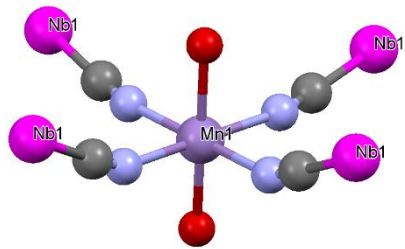

**Fig. 3:** The coordination sphere of the $Mn^{II}$ ion comprises four nitrogen atoms from the cyanido linkages and two oxygen atoms from the water molecules arranged in the slightly distorted octahedral geometry.

3. Experimental

The molar mass of the studied compound with the collective formula $NbMn_2C_8N_8O_8H_{16}$ amounts to 555.046 g mol$^{-1}$. Its density is equal to 1.89 g cm$^{-3}$. The heat capacity measurements were performed with the PPMS Quantum Design instrument by the relaxation calorimetry technique. A small amount of the compound in the polycrystalline form of mass 2.8766 mg was pressed to form a small pellet which was next attached to the sample holder by using a blob of Apiezon N grease. Prior to carrying out the sample measurements the heat capacity of the empty sample holder with the blob was scanned in both zero and nonzero applied magnetic fields to determine the reference curves. The measurements were carried out in the cooling direction in the temperature range of 1.87-151.43 K without applied field and in the range of 1.87-100.98 K in the applied field of $\mu_0H$ = 0.1, 0.2, 0.5, 1, 2, 3, 4, 5, 7, and 9 T.

4. Thermal properties

4.1 Heat capacity

Symbols in Fig. 4 show the molar specific heat of the studied compound. A characteristic lambda-shaped anomaly located around 47 K is apparent indicating the presence of the second order transition to a magnetically ordered phase. Figure 5 shows the evolution of the heat capacity anomaly due to the applied magnetic field in terms of the ratio $C_p/T$. The anomaly can be seen to shift towards higher temperatures, disperse and become suppressed with the increasing applied field value. Its shift towards higher temperatures is not expected for a system with an antiferromagnetic coupling of the constituent spins with the magnetic field acting against the tendency towards antiparallel spin arrangement. However, it can be rationalized for the system under study by realizing that the magnetic field promotes alignment of the pair of the relatively higher spins of the $Mn^{II}$ ions in the formula spin unit $S_{Mn}$–$S_{Nb}$–$S_{Mn}$ (↑↓↑) effectively enhancing the coupling. To confirm this conjecture we calculated the exchange energy average in the applied field for the spin threesome $S_{Mn}$–$S_{Nb}$–$S_{Mn}$ assuming an isotropic antiferromagnetic exchange coupling $J = -10$ K and the spectroscopic factors $g_{Mn} = g_{Nb} = 2.0$. The result is shown in Fig. 6. It is apparent that the nonzero magnetic field amplifies the exchange energy at low to intermediate temperatures. At high temperatures the field effect is opposite but slight. It may be expected that the cooperativity of the exchange interactions in the extended spin network will shift the amplification area towards higher temperatures.

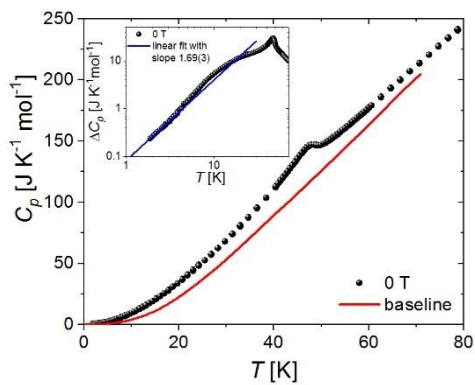

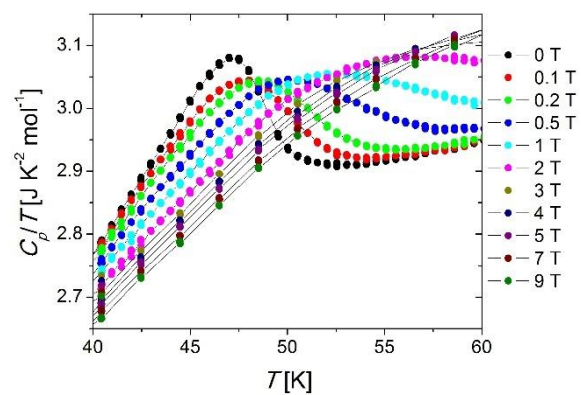

**Fig. 4:** Temperature dependence of the detected heat capacity (symbols). The solid red line shows the baseline quantifying the lattice contribution to the heat capacity. Inset: The low-temperature behavior of the zero-field excess heat capacity in the log-log plot. The solid blue line shows the linear fit yielding exponent $\mu = 1.69(3)$ (see Section 4.2)

**Fig. 5:** The evolution of the heat capacity anomaly in the applied magnetic field presented in terms of the ratio $C_p/T$. The anomaly shifts, disperses and becomes suppressed with the increasing magnetic field.

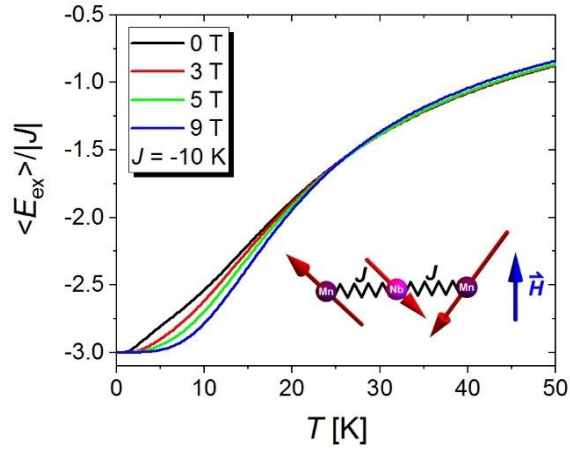

**Fig. 6:** The temperature dependence of the exchange energy average in the applied field for the spin unit $S_{Mn}$–$S_{Nb}$–$S_{Mn}$. At low and intermediate temperatures the exchange energy is enhanced by the nonzero magnetic field. At high temperatures the zero and nonzero field curves almost coincide.

4.2 Constructing the baseline

With the aim to extract the magnetic excess heat capacity and to consistently calculate the MCE quantities one must determine the lattice contribution to the heat capacity, which will be referred to as the baseline in what follows. Let us note that the studied compound is an insulator with the spin densities well-localized around the sites of the magnetic $Mn^{II}$ and $Nb^{IV}$ ions, so the heat capacity contribution from itinerant electrons can be safely neglected. The measured heat capacity comprises therefore two components, i.e. the baseline and the magnetic component. One of the most widely adopted approaches to construct the baseline is to represent the lattice contribution as a parametrized combination of Debye and Einstein functions [117,121-125]. The Einstein temperatures may be guessed on the basis of the intramolecular vibrational spectra of the constituent species like the cyanido ligand $CN^-$ or water molecules, while the Debye temperatures should be adjusted by fitting the model to the experimental heat capacity values outside the area of the transition anomaly. Despite its firm physical footing a clear disadvantage of such an approach is the uncertainty regarding the number of Debye and Einstein components and the values of the corresponding temperatures. Therefore, we propose here an approach with a more mathematical foundation drawing on the basic fact of the continuity of the baseline as a function of temperature. It is crucially based on the full magnetic entropy content of a system. The construction procedure is divided into the following steps.

- Firstly, one should fix the temperature interval $[T_1, T_2]$ such that $T_c < T_1 < T_2 \leq T_{max}$, where $T_c$ is the transition temperature and $T_{max}$ denotes the upper limit of the measurement. In that interval one should fit to the experimental zero-field heat capacity values a function comprising a polynomial approximation of the lattice contribution $C_p(\text{lattice})$ and a term corresponding to the magnetic short-range-order contribution $C_p(\text{short-range-order})$,

$$C_p = C_p(\text{lattice}) + C_p(\text{short-range-order})$$
$$= \sum_{i=3}^{n_{max}} a_i T^i + \frac{b}{T^2} \qquad (3)$$

The lowest power of the polynomial function is assumed to be equal to 3 in accordance with the low-temperature behavior of the phonon subsystem, while its degree $n_{max}$ can be varied in the fitting procedure. Apart from the experimental points (black symbols) Fig. 7 shows the boundaries of the fitting interval and the best-fit curve (solid red line).

- Secondly, the polynomial part of the best-fit $C_p(\text{lattice})$ extrapolated down to 0 K serves as the baseline candidate (see the solid blue line in Fig. 7).
- Next, one calculates the magnetic entropy contribution in the low-temperature regime for temperatures in the interval $[0, T_{min}]$ (see the cyan area in Fig. 7). To this end one extrapolates the excess zero-field heat capacity $\Delta C_p = C_p - C_p(\text{lattice})$ assuming an algebraic temperature dependence, $\Delta C_p = fT^\mu$, by performing a linear fit in the log-log plot of $\Delta C_p$ vs. $T$ in the interval where the linear trend of experimental points is apparent (see the Inset of Fig. 4). Then, with the best-fit parameter values $f$ and $\mu$ the low-temperature magnetic entropy contribution can be readily obtained

$$\Delta S_{LT} = \int_0^{T_{min}} \Delta C_p \, d\ln T = \frac{f T_{min}^\mu}{\mu} \qquad (4)$$

- Next, one calculates the main magnetic entropy contribution in the intermediate temperature regime for temperatures in the interval $[T_{min}, T_2]$ (see the green area in Fig. 7) by logarithmically integrating the excess heat capacity, i.e.

$$\Delta S_{main} = \int_{T_{min}}^{T_2} \Delta C_p \, d\ln T \qquad (5)$$

- Next, one calculates the high-temperature contribution to the magnetic entropy by logarithmically integrating the magnetic short-range-order term in the heat capacity in temperature interval $[T_2, \infty)$ (see the pink area in Fig. 7), i.e.

$$\Delta S_{HT} = \int_{T_2}^\infty bT^{-2} d\ln T = \frac{b}{2T_2^2} \qquad (6)$$

- The full magnetic entropy associated with the transition and the baseline candidate is a sum of the three contributions $\Delta S_{trans} = \Delta S_{LT} + \Delta S_{main} + \Delta S_{HT}$ and it should be compared with the maximal entropy content of the studied system $S_M^{max}$. Therefore, now we repeat the above procedure for all different choices of the interval $[T_1, T_2]$ (there is a finite number of such intervals due to the finite number of experimental heat capacity points) registering the corresponding values of $\Delta S_{trans}$. The final result corresponding to the genuine baseline is selected by requiring that the difference $S_M^{max} - \Delta S_{trans}$ is positive and minimal (see the solid red line in Fig. 4).

- Remembering that for a consistent determination of the MCE properties one needs the total entropy values $S(T,H)$ starting from the zero absolute temperature ($S(T,H) = \int_0^T C_p(T,H) \, d\ln T$), the procedure is closed by extrapolating the magnetic excess heat capacities $\Delta C_p(T,H) = C_p(T,H) - C_p(\text{lattice})$ for nonzero fields down to 0 K assuming the algebraic temperature dependence $\Delta C_p(T,H) = A_H T^{B_H}$. This is readily obtained by performing linear fits in the log-log plot of $\Delta C_p(T,H)$ vs $T$ in the intervals where the linear trend of experimental points is apparent (see Fig. 8). Then, the heat capacity in the interval $[0, T_{\min}]$ for all field values (a similar extrapolation has been performed in the zero field case previously) can be obtained by combining the baseline and the extrapolated excess heat capacity, i.e.

$$C_p(T,H) = C_p(\text{lattice}) + \Delta C_p(\text{extrapolated}) \qquad (7)$$

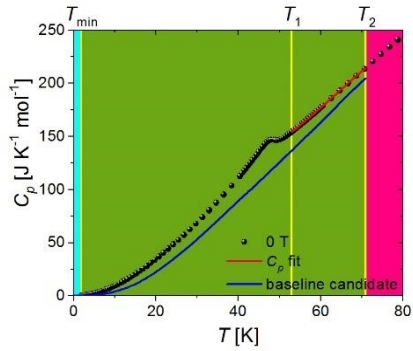
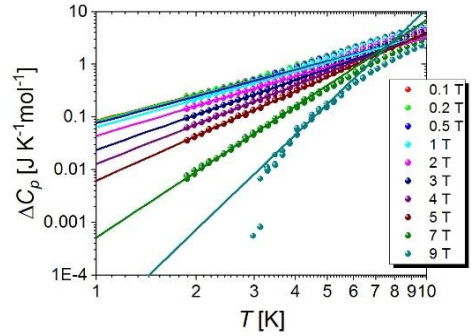

**Fig. 7:** Temperature dependence of the total heat capacity together with the best-fit curve in the interval $[T_1, T_2]$ (solid red line) and the ensuing baseline candidate (solid blue line). The colored areas (cyan, green, and pink) correspond to three disparate temperature regions where a different method to calculate a magnetic entropy contribution is employed.

**Fig. 8:** The low-temperature log-log plot of the magnetic excess heat capacities for different applied field values. A linear trend of the experimental points is apparent indicating the plausibility of the algebraic temperature dependence of $\Delta C_p(T,H)$ in the low-temperature regime.

The above procedure was successfully carried out for the compound under study. Two values of $n_{\max} = 6$ and 7 were tested. The total magnetic entropy of the compound was best reproduced by the baseline obtained with $n_{\max} = 6$, $T_1 = 53$ K, and $T_2 = 71$ K. In Table 2 the best-fit parameters and the individual magnetic entropy contributions have been collected. The corresponding value of $\Delta S_{\text{trans}} = 35.29$ J K$^{-1}$mol$^{-1}$ differs from $S_M^{\max} = 35.56$ J K$^{-1}$mol$^{-1}$ by 0.8 %. The value of zero-field exponent $\mu = 1.69(3)$ governing the low-temperature behavior of the excess heat capacity must be compared against predictions for the low-temperature heat capacity contribution of spin wave excitations displaying the $T^{d/n}$ dependence, where $d$ is the dimensionality of the spin lattice and $n$ is the exponent in the magnonic dispersion relation. It is closer to the value of 3/2, corresponding to the presence of gapless acoustic magnons with $n$

= 2 for a 3D spin network ($d$ = 3) like in the $T^{3/2}$- Bloch law valid for a 3D ferromagnet, than to that of 3 characteristic for a 3D antiferromagnet with $n$ =1. A similar behavior was reported for ferrimagnets in bipartite lattice [126-128], where the spectrum of the elementary magnetic excitations was demonstrated to split into two branches, i.e. one acoustic branch starting off at zero energy (gapless magnons) and one optical branch with its energy remaining finite for all values of the wave vector (gapped magnons). Understandably, the low-temperature behavior of the heat capacity is determined by the branch characterized by the lowest excitation energies, i.e. the gapless acoustic magnons exhibiting a quadratic (ferromagnetic) dispersion relation in the long-wavelength limit.

Figure 8 shows the log-log plot of the magnetic excess heat capacities for all applied field values. Apart from the highest field of 9 T a linear trend of the experimental points is apparent confirming the plausibility of the algebraic function governing the low-temperature behavior of the magnetic heat capacity contribution. At 9 T a significant discrepancy is observed. One the one hand, it might have been caused by some instrumental failure with two points visibly deviating from the main trend. On the other hand, the absolute heat capacity values for that field are the lowest, of the order of $10^{-2}$ J K$^{-1}$mol$^{-1}$, and a possible error incurred by the application of the algebraic law will be negligibly small.

**Table 2:** Parameters and quantities corresponding to the baseline construction for the compound under study.

| | | | |
|---|---|---|---|
| $a_3$ [J K$^{-4}$mol$^{-1}$] | $5.47(9) \times 10^{-3}$ | $f$ [J K$^{-(1+\mu)}$mol$^{-1}$] | 0.081(2) |
| $a_4$ [J K$^{-5}$mol$^{-1}$] | $-1.72(4) \times 10^{-4}$ | $\mu$ | 1.69(3) |
| $a_5$ [J K$^{-6}$mol$^{-1}$] | $2.13(7) \times 10^{-6}$ | $\Delta S_{LT}$[J K$^{-1}$mol$^{-1}$] | 0.14 |
| $a_6$ [J K$^{-7}$mol$^{-1}$] | $-9.6(4) \times 10^{-9}$ | $\Delta S_{main}$[J K$^{-1}$mol$^{-1}$] | 30.26 |
| $b$ [J K mol$^{-1}$] | 49323 | $\Delta S_{HT}$[J K$^{-1}$mol$^{-1}$] | 4.89 |

4.3 Magnetic excess heat capacity

The temperature dependence of the magnetic excess heat capacity $\Delta C_p$ for all applied field values is depicted in Fig. 9. It was obtained by subtracting the lattice heat capacity (the baseline) constructed in the previous Section from the detected heat capacity values. The position of the maximum of the zero-field anomaly marks the second order transition to the ferrimagnetic phase and is equal to 47.0(2) K, where the error is determined by the distance from the neighboring experimental points. The suppression of the anomaly and its shift towards higher temperatures with increasing magnetic field is even more clearly visible in this plot.

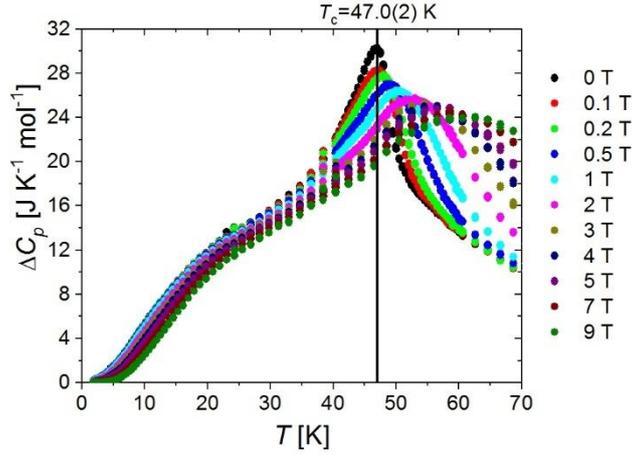

**Fig. 9:** The magnetic excess heat capacity for all applied field values obtained with the baseline constructed in Section 4.2. It is clearly visible that the anomaly peak is suppressed and shifted towards higher temperatures with increasing magnetic field.

5. Magnetocaloric effect

5.1 Isothermal change of magnetic entropy

The calorimetric method of determining the isothermal entropy change due to the field change $\mu_0 \Delta H = H_f - H_i$ is more direct than the magnetometric one using the relation in Eq. (1). Namely, it is enough to calculate the total entropies corresponding to the field values $H_i$ and $H_f$ and subtract them from each other, i.e.

$$\Delta S_M (T, H = H_i \to H_f) = \int_0^T \frac{C_p(T', H_f) - C_p(T', H_i)}{T'} dT' \qquad (8)$$

Figure 10 shows the temperature dependence of $\Delta S_M$ for $\mu_0 H_i = 0\ T$ and an array of the final field values $\mu_0 H_f =$ 0.1, 0.2, 0.5, 1, 2, 3, 4, 5, 7, and 9 T. Such a choice of the boundary values, in the normal situation corresponding to the expected inequality $S(T, H \neq 0) < S(T, H = 0)$, makes the quantity of $\Delta S_M$ negative (this case is usually referred to as the direct MCE), so the plot in Fig 10 depicts the $-\Delta S_M$ vs. $T$ dependence. In fact, the inverse MCE with $\Delta S_M > 0$ is also observed for the studied compound, but this will be discussed later. The magnetic entropy change curves in Fig. 10 can be seen to display a peak slightly above the transition temperature. Moreover, there are broad hump-like structures on the left shoulders of the peaks. The behavior of the peak values (red symbols) and positions (blue symbols) as a function of the applied field change can be traced in Fig. 11. Both the peak amplitudes and the peak positions show an increasing trend with increasing magnitude of the magnetic field change. While the peak amplitude curve is continuous, the peak position behavior is more abrupt probably due to the experimental distribution of the temperature points. The fact that the $|\Delta S_M|$ peaks are located slightly above the transition temperature and tend to shift towards higher temperatures with the increasing applied field-change values may be demonstrated to be related with the analogous shift of the magnetic anomaly with increasing fields discussed in Section 4.1 and apparent from

Fig. 5. It is enough to look at the necessary condition for the extremum of $\Delta S_M(T, 0 \to H)$ as a function of temperature (see Eq. (8)), i.e.

$$\frac{\partial \Delta S_M(T, 0 \to H)}{\partial T} = \frac{C_p(T, H) - C_p(T, 0)}{T} = 0,$$

which immediately implies that $C_p(T_{peak}, H) = C_p(T_{peak}, 0)$. This means that the peaks are placed exactly there where the heat capacity curve in a nonzero magnetic field intersects that corresponding to the zero field. Upon smoothing the experimental $C_p$ data the last equation was solved for all magnetic field values. The solid blue line in Fig. 11 shows the result. It is consistent with the experimental values of $T_{peak}$. The tiny kink present at the lowest field value is probably due to the smoothing procedure and the sparse sampling of the data does not allow its verification.

The yellow stars in Fig. 10 show the $\Delta S_M$ values obtained previously with the magnetometric method [115]. The agreement is satisfactory, however the previous values are systematically slightly underestimated on the left shoulders and simultaneously slightly overestimated on the right shoulders of the peaks. Also the peak values seem to be slightly shifted towards higher temperatures. One the one hand, these discrepancies might have been caused by a systematic overestimation of the measured temperature by the applied magnetometer or a systematic underestimation of the sample temperature detected by the PPMS instrument. Indeed, it would be sufficient to slightly shift the yellow-star curves towards lower temperatures to make both signal perfectly coincide. On the other hand, the same effect would be produced by the calculation mode of the derivative $\partial M(T, H)/\partial T$ in Eq. (1) with the difference quotient $[M(T_{i+1}, H) - M(T_i, H)]/(T_{i+1} - T_i)$ associated with the interval $[T_i, T_{i+1}]$ assigned to its right boundary instead of to its center. With the discrete temperature sampling enhancing this effect we think the latter is the case.

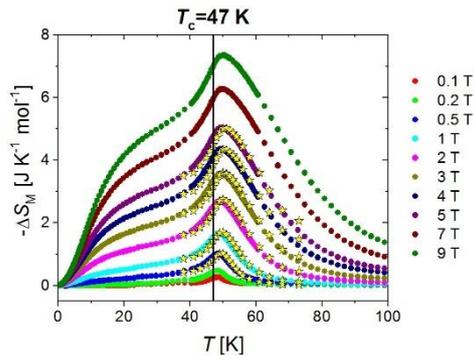 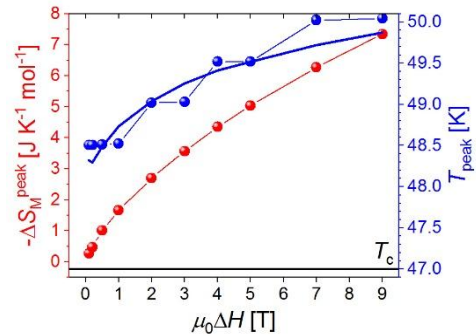

**Fig. 10:** Temperature dependence of the isothermal entropy change $\Delta S_M$ for indicated magnetic field changes. Yellow stars show the $\Delta S_M$ values obtained previously with the magnetometric method [115].

**Fig. 11:** Peak values of $-\Delta S_M$ together with the peak positions. The solid blue line shows the theoretical prediction (see main text).

5.2 Modelling $\Delta S_M$ – mean field approach

We use a molecular field approximation in order to get a more intuitive insight into the origins of the magnetic entropy changes in the studied compound. The first step of constructing the corresponding model is the analysis of its crystal structure. We know from Section 2 that it consists of two sublattices: one corresponding to the two crystallographically equivalent $Mn^{II}$ ions (there will be a double occupancy of this sublattice) and the other one to the $Nb^{IV}$ ions (with a single occupancy). Apart from the Zeeman coupling of the spins with the applied magnetic field we take into account the superexchange coupling between the nearest neighbor ions assuming a single coupling constant $J_{MnNb}$ which obeys the convention that its negative values correspond to antiferromagnetic interaction. The main component of the model is a set of coupled equations expressing the thermal averages of the sublattice spins [129]:

$$\langle S_{Mn} \rangle = S_{Mn} B_{S_{Mn}} \left( \frac{g_{Mn} \mu_B S_{Mn} H}{k_B T} + \frac{Z_{MnNb} J_{MnNb} S_{Mn}}{k_B T} \langle S_{Nb} \rangle \right)$$
$$\langle S_{Nb} \rangle = S_{Nb} B_{S_{Nb}} \left( \frac{g_{Nb} \mu_B S_{Nb} H}{k_B T} + \frac{Z_{NbMn} J_{MnNb} S_{Nb}}{k_B T} \langle S_{Mn} \rangle \right)$$
, (9)

where $S_X$ (X = Mn, Nb) is the value of $\langle S_X \rangle$ at $T = 0$ K, $B_S$ is the spin-$S$ Brillouin function, $g_X$ (X = Mn, Nb) is the spectroscopic factor, $\mu_B$ denotes the Bohr magneton, $H$ is the external magnetic field, $Z_{XY}$ (X, Y = Mn, Nb) are the numbers of the nearest neighbor are the numbers of the nearest neighbor Y-site centers surrounding an X-site center, and $k_B$ is the Boltzmann constant. Numerically iterating the set in Eq. (9) yields the mean values $\langle S_X \rangle$ and the molar total magnetization is calculated as:

$$M = N_A \mu_B [\lambda_{Mn} g_{Mn} \langle S_{Mn} \rangle + \lambda_{Nb} g_{Nb} \langle S_{Nb} \rangle], \quad (10)$$

where $N_A$ is the Avogadro number, and $\lambda_X$ (X = Mn, Nb) represents an occupancy factor of center X. A close inspection of Fig. 2 and 3 in Section 2 yields $Z_{MnNb} = 4$ and $Z_{NbMn} = 8$. The occupancy factors corresponding to the compound under study are $\lambda_{Mn} = 2$ and $\lambda_{Nb} = 1$. The spectroscopic factors were fixed during the calculations at the isotropic values of $g_{Mn} = g_{Nb} = 2.0$, and the corresponding spin values are $S_{Mn} = 5/2$ and $S_{Nb} = \frac{1}{2}$. The simulation consists in the generation of a map of magnetization values for a discrete distribution of temperature and magnetic field variables and next using formula in Eq. (1) to evaluate the isothermal magnetic entropy change $\Delta S_M$ corresponding to the applied field change $H = 0 \to H_f$ with $\mu_0 H_f \in [0,9]$ T. However, prior to simulating MCE one must adjust the value of the superexchange coupling constant $J_{MnNb}$, so that it reproduces the magnetic properties of the studied compound. Apart from the total magnetization and sublattice magnetizations simulated within the molecular approximation in zero applied field Fig. 12 shows the temperature dependence of the corresponding derivative $dM_{total}/dT$ (red) obtained assuming the coupling constant to be equal to $J_{MnNb} = -10.26$ K. The apparent peak marks the position of the second-order phase transition coinciding with $T_c = 47$ K found experimentally (see Section 4.3). Let us note that the tuned value of the coupling constant $J_{MnNb}$ sets the lower limit for its true value,

as the mean-field approximation tends to overestimate the transition temperature for a given exchange coupling value.

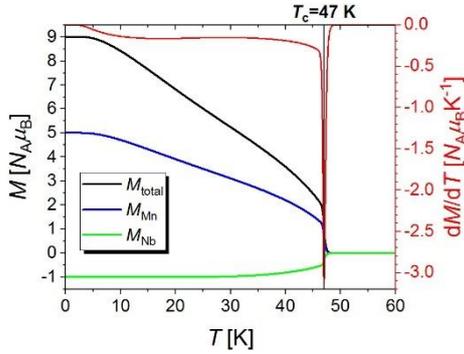 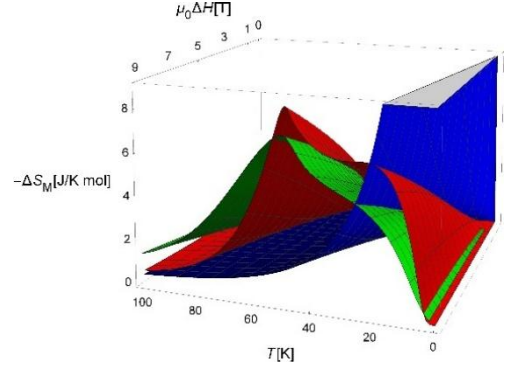

**Fig. 12:** Left axis: The temperature dependence of the total magnetization (black) together with Mn (blue) and Nb (green) sublattice magnetizations in zero applied field calculated within the molecular field approximation with $J_{\text{MnNb}} = -10.26$ K. Right axis: The temperature dependence of the corresponding temperature derivative $dM_{\text{total}}/dT$ in zero applied field (red). The peak marks the position of the magnetic order-disorder transition.

**Fig. 13:** The temperature and field dependence of $\Delta S_{\text{M}}$ for the studied compound in three cases: as obtained by the calorimetric method (green), simulated within the molecular approximation with $J_{\text{MnNb}} = -10.26$ K (red), and calculated for a system of isolated $\text{Mn}^{\text{II}}$ and $\text{Nb}^{\text{IV}}$ ions with $J_{\text{MnNb}} = 0$ (blue). The blue surface significantly exceeds the plot scale reaching the value of $S_{\text{M}}^{\max}$.

Figure 13 shows the temperature and field dependence of the isothermal magnetic entropy change $\Delta S_{\text{M}}$ due to the switching on the field ($\Delta H = H_{\text{f}} - H_{\text{i}}(=0) > 0$) for three cases: the quantity obtained by the calorimetric method (the green surface), simulated within the molecular approximation approach with $J_{\text{MnNb}} = -10.26$ K (the red surface), and for a paramagnetic system of decoupled $\text{Mn}^{\text{II}}$ and $\text{Nb}^{\text{IV}}$ ions ($J_{\text{MnNb}} = 0$, the blue surface). The mean-field model can be seen to give smaller $|\Delta S_{\text{M}}|$ values at higher temperatures above the transition point. This observation can be understood by remembering that there is an intrinsic lack of spin correlations (fluctuations) in the molecular field approximation. These fluctuations tend to diminish the magnetic entropy in nonzero field above the transition due to the pretransitional short-range order effects resulting in a more pronounced entropy difference $\Delta S_{\text{M}}$. Indeed, a fluctuating pattern of locally ordered islands of the correlation-length size is more susceptible to alignment by the applied magnetic field as compared to the perfectly disordered purely paramagnetic state. At the same time, the opposite behavior is observed below the transition points with the mean-field values of $|\Delta S_{\text{M}}|$ being larger than the experimental ones. However, the same factor seems to play a role also in this case. As a matter of fact, the ordered state in the real compound is more difficult to be reorganized by the applied magnetic field as consisting of misaligned fluctuating domains, while that in the mean-field model is a robust monodomain structure. Hence, the magnetic entropy associated with that single domain of the mean-field model is understandably lower than that corresponding to the multidomain system

of the real compound, which makes the experimental value of the entropy change $|\Delta S_M|$ systematically inferior. The situation may be reversed at the lowest temperatures, where thermal fluctuations are substantially weaker (see the plot in Fig. 13 close to the 0 K edge). Finally, the entropy change for the purely paramagnetic state corresponding to the magnetic decoupling of the Mn$^{II}$ and Nb$^{IV}$ ions (the blue surface) was included for the sake of comparison. The plots in Fig. 13 clearly demonstrate how, due to the exchange coupling, the magnetic degrees of freedom are shifted from the absolute zero temperature. This occurs at the apparent price of the entropy change amplitude $|\Delta S_M|$ with its maximal value reaching the physical limit of $S_M^{max}$ in the paramagnetic state (not shown in Fig. 13 for the sake of clarity) and assuming the value roughly four times smaller for the coupled case. The position of the $|\Delta S_M|$ peak depends on the exchange coupling strength $J_{MnNb}$, so does its amplitude through the elementary relation between entropy and enthalpy $\delta S = \delta Q / T$, where $\delta Q$ quantifies the energy of exchange interactions.

5.3 Adiabatic temperature change

With entropy being a monotonically increasing function of temperature (indeed, $(\partial S / \partial T)_H = C_p / T > 0$) one can calculate the adiabatic change of temperature associated with the field change $H = H_i \to H_f$ by inverting the temperature dependence of the entropy state function, i.e.

$$\Delta T_{ad}(T, \Delta H = H_f - H_i) = [T(S, H_f) - T(S, H_i)]_S. \tag{11}$$

In a normal situation, switching on magnetic field adiabatically ($H_i = 0$, $H_f \neq 0$) causes a sample to heat up, i.e. $\Delta T_{ad} > 0$ and one refers to it as the direct MCE. Should a sample cool down while adiabatically increasing the field, i.e. $\Delta T_{ad} < 0$, then the phenomenon is called the inverse MCE. Just the opposite holds for the adiabatic demagnetization of a sample ($H_i \neq 0$, $H_f = 0$), i.e. $\Delta T_{ad} < 0$ for the direct MCE and $\Delta T_{ad} > 0$ for the inverse one. Figure 14 shows the temperature dependence of $\Delta T_{ad}$ due to switching on magnetic field (symbols, magnetizing the sample) as well as due to switching off the field (solid orange lines, demagnetizing the sample) obtained for the studied compound by using the recipe in Eq. (11). Both families of curves corresponding to the different applied field change values display a two-peak structure with the demagnetization curves shifted slightly towards higher temperatures. The primary peaks are located slightly above the transition temperature and their amplitudes are monotonically increasing with the magnetic field change. The behavior of the peak values (red symbols) and positions (blue symbols) for the magnetization process as a function of the applied field change can be traced in Fig. 15. A comparison of the $T_{peak}$ data in Fig. 11 with that in Fig. 15 (blue symbols) allows one to conclude that the peak temperatures of the isothermal magnetic entropy change differ from those of the adiabatic temperature change. This fact can be explained by considering the necessary condition for the primary peaks of $\Delta T_{ad}$. Differentiating Eq. (11) with respect to the temperature variable and equating the result to zero one arrives at the following implicit equation for $T_{peak}$:

$$(T_{peak} + \Delta T_{ad}^{peak}) C_p (T_{peak}, H_i) = T_{peak} C_p (T_{peak} + \Delta T_{ad}^{peak}, H_f). \tag{12}$$

It can be approximately solved within a dedicated perturbation scheme, where one expands the solution in powers of the small parameter $\varepsilon = |\Delta T_{ad}^{peak}|/T_{peak}$. The final solution may be written as

$$T_{peak} = T_{peak}^{(0)} + f_{M/DM} |\Delta T_{ad}^{peak}|, \qquad (13)$$

where the zeroth order term $T_{peak}^{(0)}$ satisfies the same equation as $T_{peak}$ for $\Delta S_M$, i.e.

$$C_p(T_{peak}^{(0)}, H) = C_p(T_{peak}^{(0)}, 0), \qquad (14)$$

And the prefactors $f_{M/DM}$ corresponding to the magnetization ($H_i = 0, H_f = H$) and demagnetization ($H_i = H, H_f = 0$) process, respectively, are given by the formulas:

$$f_{M/DM} = \frac{\dfrac{C_p(T_{peak}^{(0)}, 0/H)}{T_{peak}^{(0)}} - \dfrac{\partial C_p}{\partial T}(T_{peak}^{(0)}, H/0)}{\dfrac{\partial C_p}{\partial T}(T_{peak}^{(0)}, H) - \dfrac{\partial C_p}{\partial T}(T_{peak}^{(0)}, 0)}, \qquad (15)$$

Using smoothed $C_p$ data the values of both prefactors were calculated. Their field change dependence is shown in Inset of Fig. 15. It can be seen that for the demagnetization process the prefactors are always positive ($f_{DM} > 0$), while for the magnetization process they are negative ($f_M < 0$) except for one point at the lowest field change value of 0.1 T. This departure from the dominant trend may be an artifact of the smoothing procedure of the sparse experimental data and cannot be easily verified. The signs of the prefactors mean that the peak temperatures of $\Delta T_{ad}$ shift slightly off the peak temperatures of $\Delta S_M$ to the left for the magnetization process and to the right for the demagnetization process. This is consistent with the relative shift of the magnetization and demagnetization $\Delta T_{ad}$ curves in Fig. 14. Moreover, the solid blue line in Fig. 15 showing the values of $T_{peak}$ for the magnetization process calculated using Eqs. (13-15) agrees well with the experimental points.

The secondary peaks are broad and spread from low to intermediate temperatures. Their amplitudes monotonically increase with the magnetic field change exceeding the primary peaks for the two highest field change values. A possible reason for their appearance may be seen by inspecting the formula in Eq. (2). Indeed, from Fig. 10 it may be concluded that for low to intermediate temperatures the temperature dependence of the isothermal entropy change is roughly linear, i.e. $\delta S_M(T, \delta H) \sim T$, while the dominating temperature power in the heat capacity is 3, i.e. $C_p(T, H) \sim T^3$. Then, according to Eq. (2) the amplitude of the adiabatic temperature change is expected to behave like $\Delta T_{ad} \sim \dfrac{T \cdot T}{T^3} = T^{-1}$ in this temperature regime leading to its observed systematic increase with lowering the temperature.

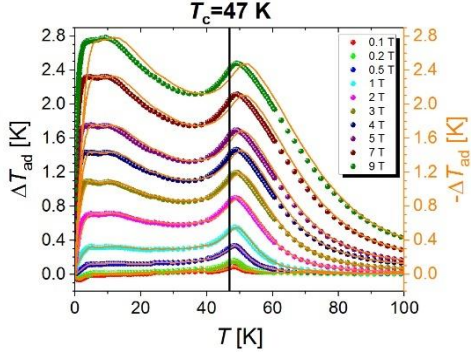
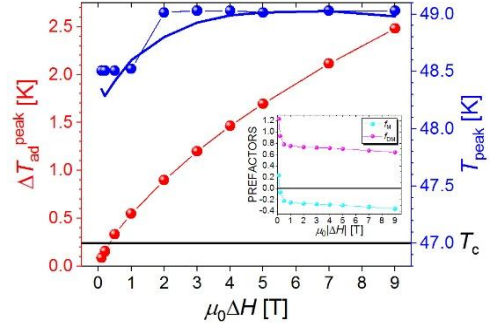

**Fig. 14:** The temperature dependence of the adiabatic change of temperature $\Delta T_{ad}$ associated with magnetization (symbols) and demagnetization (solid orange lines) of the sample under study.

**Fig. 15:** Peak values of $\Delta T_{ad}$ together with peak positions for magnetization of the studied compound. The solid blue line is the theoretical prediction (see main text). Inset: The applied field change dependence of the prefactors used to calculate $T_{peak}$ within the perturbation scheme for magnetization ($f_M$) and demagnetization ($f_{DM}$) processes (see main text).

5.4 The inverse MCE

A close inspection of the temperature dependences of $\Delta S_M$ and $\Delta T_{ad}$ (see Fig. 10 and 14) shows that both MCE quantities change sign deeply below the transition temperature for the three lowest field change values, i.e. $\mu_0 \Delta H = 0.1$, 0.2, and 0.5 T. It is apparent in Fig. 16, where the $\Delta S_M$ vs. $T$ and $\Delta T_{ad}$ vs. $T$ functions for $\mu_0 \Delta H = 0.1$ and 0.2 T are depicted. For the field change of 0.5 T the inverse magnetocaloric effect is marginal but still present. The inset of Fig. 16 shows the close-up of the low temperature regime, where we can see the simultaneous onset of the inverse MCE for both quantities. The occurrence of the inverse MCE is consistent with the low-temperature behavior of the excess heat capacities (see Section 4.2). Using the assumed algebraic functions for $\Delta C_p$ and the formulas in Eqs. (7) and (8) one can readily arrive at the following result for low temperatures ($T \to 0$):

$$\Delta S_M(T, 0 \to H) = \int_0^T \frac{\Delta C_p(T',H) - \Delta C_p(T',0)}{T'} dT' \approx \begin{cases} -\dfrac{f}{\mu} T^\mu & \text{for } B_H > \mu \\ +\dfrac{A_H}{B_H} T^{B_H} & \text{for } B_H < \mu \end{cases}. \quad (16)$$

Knowing that $B_H$ = 1.62(2), 1.64(2), and 1.68(3) for $\mu_0 H = 0.1$, 0.2, and 0.5 T, respectively, and $\mu = 1.69(3)$, that is $B_H < \mu$, one can expect a positive amplitude of $\Delta S_M$, i.e. the inverse MCE, at low temperatures. A similar behavior was reported for two molecular magnets based either likewise on the Mn[II] and Nb[IV] ions or the Fe[II] and Nb[IV] ions and the pyrazole ligand [111,117], but the inverse MCE was observed there below as well as above the transition temperature. The common feature with the compound under study is the appearance of the

inverse effect for the lowest field change values and the presence of an antiferromagnetic coupling between the constituent ions. Hence, a possible explanation could be that the relatively low field is unable to reorient the magnetic domains and the only effect it can cause is the local flips (alignment) of the antiferromagnetically coupled neighboring spins within the domains. This can be seen as a disordering factor which increases the entropy value, so that the entropy in nonzero field becomes slightly higher than that in zero field.

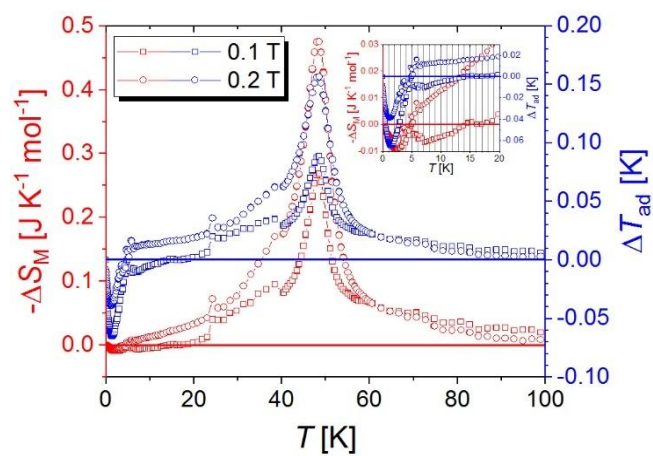

**Fig. 16:** The temperature dependence of $\Delta S_M$ (red symbols, the left axis) and $\Delta T_{ad}$ (blue symbols, the right axis) for the two lowest applied field change values. Inset: The close-up of the low-temperature regime showing the simultaneity of the onset of the inverse MCE for both quantities.

5.5 Comparisons

Cyanido-bridged magnetic coordination polymers exhibiting a second-order phase transition constitute a natural reference point for the magnetocaloric performance of the compound under study. Examples of such systems have been collected in Table 1. Upon looking at the magnitudes in Table 1 and the curves marked in red symbols in Fig. 11 and 15 one can conclude that the studied compound displays typical values for $\Delta S_M$ and $\Delta T_{ad}$. A closer inspection of the items in Table 1 shows that there are three compounds that are most suitable for a comparative analysis because they share the same magnetic ions in the same molar proportion, namely compounds in position No. 1, 5, and 8. Besides, all the three compounds exhibit likewise a ferrimagnetic phase. The main difference in their composition is that unlike the studied compound they contain an additional organic ligand coordinating the $Mn^{II}$ ions. This is pyridazine for compound No. 1, pyrazole for compound No. 5, and imidazole for compound No. 8. Since we do not have complete calorimetric data for compound No. 1 we will carry on the comparison with the two other compounds. Let us just note that compound No. 1 is characterized by a similar transition temperature and displays similar magnitudes of the MCE quantities. For the sake of reference convenience let us denote the studied compound by **1**, the compound containing the imidazole ligand by **2**, and finally, the compound containing the pyrazole ligand by **3**. In this way, the numbering follows the decreasing values of transition temperatures with $T_c(\mathbf{1}) = 47.0$ K $> T_c(\mathbf{2}) = 24.1$ K $> T_c(\mathbf{3}) = 22.8$ K. Figure 17 shows the applied field change dependence of the molar magnetic entropy changes detected at the peaks for

compounds **1** to **3**. It is apparent that the order of the curves follows the direction of increasing transition temperature with the compound displaying the lowest $T_c$ exhibiting the highest values. This may be rationalized by considering the relation between the heat amount associated with the magnetization process $\delta Q$ and the isothermal magnetic entropy change $\delta S$, i.e. $\delta S = \delta Q / T$. With the heat $\delta Q$ being comparable for the compounds comprising the same spin suit whose coupling is mediated by the same cyanido ligands providing interaction energies of the same order of magnitude, i.e. $\delta Q(\mathbf{1}) \approx \delta Q(\mathbf{2}) \approx \delta Q(\mathbf{3})$, one is to expect the entropy changes being roughly inversely proportional to temperature, i.e. $\delta S(\mathbf{1}) < \delta S(\mathbf{2}) < \delta S(\mathbf{3})$ for the entropy changes detected in the vicinity of the corresponding transition temperatures. This conjecture is even stronger confirmed by the close-lying curves corresponding to compounds **2** and **3** showing very similar transition temperatures. The situation changes substantially if one considers the magnetic entropy changes per unit mass (entropy densities) being more interesting from the applicational point of view. This is shown in Fig. 18 where the molar masses of the three compounds have been provided. Now compound **3** characterized by the largest molar mass shows the lowest $|\Delta S_M^{peak}|$ values, with the two other compounds almost coinciding at relatively higher magnitudes. Hence, one can come to the conclusion that regarding the magnitude of the magnetothermal effect it is more beneficial to have a compound with the least possible molecular weight which together with the high-spin content would produce a large entropy density.

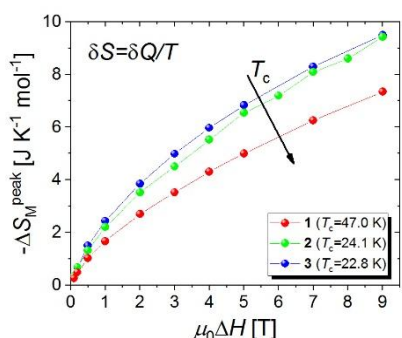
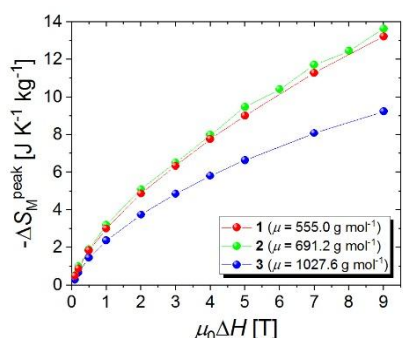

**Fig. 17:** The applied field change dependence of the molar magnetic entropy changes detected at the peaks for compounds **1** to **3** (see main text). The arrow shows the direction of increasing transition temperature.

**Fig. 18:** The applied field change dependence of the mass magnetic entropy change detected at the peaks for compounds **1** to **3** (see main text).

Figure 19 shows the magnetic field change dependence of the adiabatic temperature changes detected at the peaks for the three compounds **1** to **3**. Here the order of the curves is different with the highest $\Delta T_{ad}^{peak}$ values observed for compound **2**, intermediate ones for the studied compound **1**, and the lowest ones for compound **3**. This can be explained by combining three facts. Firstly, the adiabatic temperature change is related to the isothermal molar magnetic entropy change multiplied by the temperature to molar heat capacity ratio, which is expressed by formula in Eq. (2). Secondly, the molar magnetic entropy changes at the peaks satisfy the following inequality $\Delta S_M^{peak}(\mathbf{1}) < \Delta S_M^{peak}(\mathbf{2}) \leq \Delta S_M^{peak}(\mathbf{3})$ (cf. Fig. 17). And finally, the magnitudes of the $T/C_{p0}$ ratio observed at the transition temperatures arrange in the following order

$T/C_{p0}(\mathbf{1}) \geq T/C_{p0}(\mathbf{2}) > T/C_{p0}(\mathbf{3})$ (see Fig. 20). Let us note that $T/C_{p0}$, where the zero-field heat capacity is used, sets the lower limit for the values of the ratio $T/C_p$, where the general heat capacity in nonzero magnetic field is present. The plot in Fig. 20 implies that $C_{p0}(\mathbf{1}) < C_{p0}(\mathbf{2}) < C_{p0}(\mathbf{3})$, which reflects the order of the molecular weights, i.e. $\mu(\mathbf{1}) < \mu(\mathbf{2}) < \mu(\mathbf{3})$, and again we are made to conclude that it is low molar weight that favors high thermal effect in terms of $\Delta T_{ad}$.

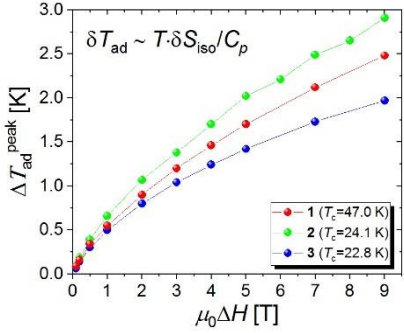

**Fig. 19:** The applied field change dependence of the adiabatic temperature changes detected at the peaks for compound **1** to **3** (see main text)

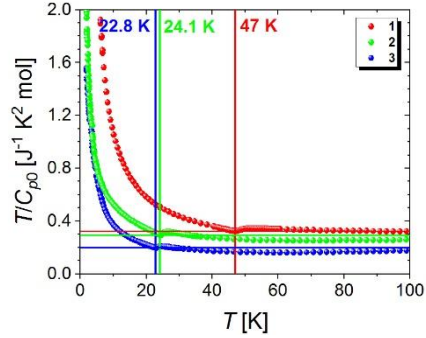

**Fig. 20:** The temperature dependence of the ratios $T/C_{p0}$ for compounds **1** to **3** (see main text). The horizontal solid lines are the guidelines for the ratio values assumed at the corresponding transition temperatures.

6. Field dependence of MCE

Theoretical analyses based on the mean-field model [130], the scaling equation of state [131] and the empirical Arrott-Noakes equation of state [132] have all led to the conclusion the field dependence of the isothermal entropy change $|\Delta S_M(T, 0 \rightarrow H)|$ for systems with the second-order phase transition displays a universal power-law behavior at the transition temperature $T_c$, i.e. $|\Delta S_M| \propto H^n$. Moreover, it was found that there is a relationship between exponent $n$ at $T_c$ and the critical exponents of a material, i.e.

$$n\big|_{T=T_c} = 1 + \frac{\beta - 1}{\beta + \gamma}. \tag{17}$$

So far in our research we have concentrated on deriving the field averaged values of exponent $n$ [105,108,109,111,115,117,119]. Here we present a more detailed analysis in a wide temperature range encompassing the transition point. Figure 21 shows the temperature and field dependence of exponent $n$ for the compound under study obtained using the second-order interpolation and the three-point approximants for the field derivatives. At the lowest temperatures exponent $n$ assumes the values close to and below 1 depending on the field value. For the fields of 1 and 2 T the curves start off at higher values and are decreasing with increasing temperature to reach a minimum somewhere close to transition temperature. For the higher fields above 2 T the curves start off at lower values, are first increasing and then decreasing with increasing temperature, and attain a minimum close to the transition point. Above the minimum all curves are increasing functions of temperatures up to about 65 K, beyond which

they show a very irregular behavior accompanied with large numerical errors attaining values close to 2. The solid black line shows the averaged values of exponent *n*. Its course is similar to that of the curves corresponding to the higher field values. The data obtained previously with the magnetometric method [115], shown by the yellow stars, coincide with the curves for the low field values. This suggests that the low field $\Delta S_M$ vs. $H$ dependence was used to derive them. This conjecture would also be consistent with the discrepancy observed between them and the averaged values of *n* (cf. the solid black line in Fig. 21) calculated on the basis of the wide field interval from 1 to 9 T. The Insert of Figure 21 shows the close-up of the vicinity of the transition temperature. It is apparent that at no temperature point the value of exponent *n* is universal. Indeed, it displays a dispersion of values at $T_c$ as well as at the local minima whose positions change with the applied field (cf. the solid red line in Insert of Fig. 21 being the eyeguide following the local minima of *n* vs. *T* dependence). The positions of the local minima may be seen to be shifted off $T_c$ towards higher temperatures. At the same time they are located at the left border of the interval for $T_{peak}$ corresponding to $\Delta S_M^{peak}$ (cf. the yellowish area in Insert of Fig. 21).

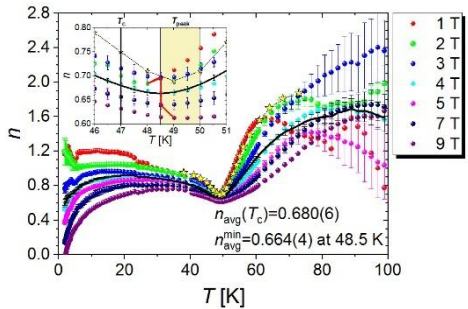
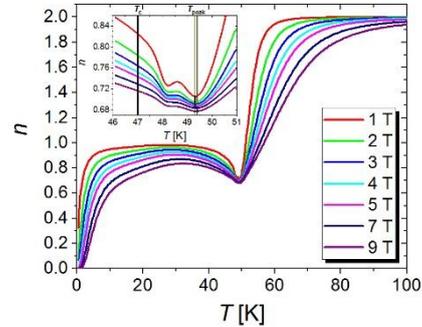

**Fig. 21:** The temperature dependence of exponent *n* for an array of applied field values for the studied compound. The solid black line shows the averaged data. The yellow stars correspond to the data obtained previously with the magnetometric method [115]. Insert: The close-up of the vicinity of the transition temperature. The yellowish area shows the range of variation of $T_{peak}$ for $\Delta S_M^{peak}$. The solid red line is the eyeguide following the local minima of *n* for the different field values.

**Fig. 22:** The temperature dependence of exponent *n* for an array of applied field values derived within the mean-field model developed in Section 5.2. Insert: The close-up of the vicinity of the transition temperature. The narrow yellowish area shows the range of variation of $T_{peak}$ for $\Delta S_M^{peak}$. The solid black line is the eyeguide following the global minima of *n* for the different field values.

For the sake of comparison the temperature and field dependence of exponent *n* was calculated within the mean-field model developed in Section 5.2. The result is shown in Fig. 22. The curves may be seen to display a regular behavior for all applied field values. At the lowest temperature they start off from the values close to 0, next they are first increasing and then decreasing with increasing temperature, next they attain a minimum, above which they increase to saturate at the value of 2 at high temperatures. The saturation value is consistent with the validity of the Curie-Weiss law in this temperature regime, where magnetization is proportional to the applied field, i.e. $M = \chi_{C-W} H$ and the formula in Eq. (1) immediately implies $n = 2$. The

Insert of Fig. 22, showing the close-up of the vicinity of the transition temperature, reveals that at least for the low field values there are two local minima both shifted off $T_c$ towards higher temperatures. The global minima (cf. the solid black line being the eyeguide following their positions) are located within the narrow interval of $T_{peak}$ corresponding to $\Delta S_M^{peak}$ (see the yellowish stripe). Similarly, at no temperature exponent $n$ shows a universal behavior, that is independent of the applied field value. The temperature dependence of $n$ for the studied compound may be concluded to be roughly consistent with the mean-field prediction with the most striking difference being the presence of a single minimum of the experimental curves instead of two local minima of the mean-field ones. Let us also note the dramatic difference in the width of the $T_{peak}$ intervals (the yellowish areas in Inserts of Fig. 21 and 22), but this may be due to the relatively low temperature resolution of the experimental data with the increments amounting to 0.5 K.

Finally, let us discuss the reliability of the critical scaling predicted by the theoretical models (see formula in Eq. (17)). The value of $n$ detected at the transition temperature for applied fields ranging from 1 to 9 T displays a clear dispersion (see Insert of Fig. 21) with the quite wide range of variation between 0.632 and 0.716 and the average value of $n_{avg}(T_c) = 0.680(6)$. One should compare these values of $n$ with those predicted by Eq. (17) for the typical models of the three-dimensional (3D) spin networks such as the mean-field model, the Heisenberg model and the Ising model. The corresponding values of the $\beta$ and $\gamma$ exponents together with the values of exponent $n$ calculated using the formula in Eq. (17) have been collected in Table 3 for the sake of a handy reference. It can be seen that the experimental range of $n$ detected at $T_c$ encompasses two out of three models, i.e. the mean-field model and the Heisenberg model, with the averaged value of $n$ being closer to the mean-field one. This observation reflecting the fact that neither the experiment (see Insert of Fig. 21) nor the molecular-field calculation (see Insert of Fig. 22) confirms the universal behavior (field-independence) of exponent $n$ at $T_c$ makes a reliable conclusion concerning the actual universality class very difficult if not impossible.

**Table 3:** Critical exponents for selected 3D models

| Model | $\beta$ | $\gamma$ | $n$ (cf. Eq. (17)) |
|---|---|---|---|
| Mean-field | 0.5 | 1.0 | $0.(6) \approx 0.67$ |
| Heisenberg [133] | 0.3689(3) | 1.3960(9) | 0.6424(4) |
| Ising [134] | 0.3265(3) | 1.2372(5) | 0.5693(4) |

7. Analysis of thermodynamic cycles

The thermodynamic cycles of particular technological importance within the context of the magnetic refrigeration are the regenerative Brayton cycle and the regenerative Ericsson cycle [64,135-138]. In this Section we shall discuss performance of the studied compound when used as a working substance in these cycles.

7.1 The regenerative Brayton cycle

First let us consider the regenerative Brayton refrigeration cycle $A \rightarrow B \rightarrow D \rightarrow E \rightarrow A$ whose scheme in the ($T$, $S$) plane is shown in Fig. 23. It consists of two adiabatic processes ($A \rightarrow B$ starting at $T_H$ and $H = 0$, and $D \rightarrow E$ starting at $T_C$ and $H \neq 0$) and two isofield processes (

$B \to D$ at applied field $H$, and $E \to A$ at zero applied field). $T_H$ and $T_C$ denote temperatures of the hot and cold reservoirs, respectively. The isofield process $B \to D$ is split into two subprocesses $B \to C$ and $C \to D$ characterized by the following heat amounts,

$$Q_H = \int_{B \to C} TdS = \int_{T_H+\Delta T_H}^{T_H} T\frac{\partial S(T,H)}{\partial T}dT = -\int_{T_H}^{T_H+\Delta T_H} C_p(T,H)dT < 0,$$

$$Q_{SR} = \int_{C \to D} TdS = \int_{T_H}^{T_C} T\frac{\partial S(T,H)}{\partial T}dT = -\int_{T_C}^{T_H} C_p(T,H)dT < 0, \quad (18)$$

where $\Delta T_H > 0$ is the adiabatic temperature change in the process $A \to B$. $Q_H$ and $Q_{SR}$ are both negative, which means that they quantify the heats released from the working substance to the hot reservoir and the regenerator, respectively. Similarly, the zero-field process $E \to A$ is split into two subprocesses $E \to F$ and $F \to A$. The corresponding heat quantities are given by the following formulas:

$$Q_C = \int_{E \to F} TdS = \int_{T_C-\Delta T_C}^{T_C} T\frac{\partial S(T,0)}{\partial T}dT = \int_{T_C-\Delta T_C}^{T_C} C_p(T,0)dT > 0,$$

$$Q_{RS} = \int_{F \to A} TdS = \int_{T_C}^{T_H} T\frac{\partial S(T,0)}{\partial T}dT = \int_{T_C}^{T_H} C_p(T,0)dT > 0, \quad (19)$$

where $-\Delta T_C < 0$ denotes the adiabatic temperature change in the process $D \to E$. $Q_C$ and $Q_{RS}$ are both positive, which means that they quantify the heats transferred to the working substance from the cold reservoir and the regenerator, respectively. The quantity affecting the cooling efficiency of the regenerative Brayton refrigeration cycle is the redundant or inadequate regenerative heat defined as

$$\Delta Q = -(Q_{SR} + Q_{RS}) = \int_{T_C}^{T_H} [C_p(T,H) - C_p(T,0)]dT. \quad (20)$$

If $\Delta Q > 0$, i.e. if the heat transferred from the working substance to the regenerator exceeds that transferred from the regenerator to the working substance, then that surplus of heat must be released to the cold reservoir, otherwise it would lead to the temperature change of the regenerator which will stop operating properly. In this case the net cooling quantity is equal to $Q_L = Q_C - \Delta Q$. If the opposite holds, i.e. $\Delta Q < 0$ and the heat transferred from the working substance to the regenerator is smaller than the heat absorbed from the regenerator by the working substance, the inadequate regenerative heat can be compensated from the hot reservoir and the cooled space (the cold reservoir) remains unaffected. Thus, the net cooling quantity is then $Q_L = Q_C$. To proceed with the discussion let us make the following observation: function $-\Delta S_M(T, 0 \to H)$ (note the minus sign in front of the quantity) is an increasing function of temperature for $T < T_{peak}$ and a decreasing function of temperature for $T > T_{peak}$ (cf. Fig. 10); the formula in Eq. (8) implies that

$$-\frac{\partial \Delta S_M(T, 0 \to H)}{\partial T} = \frac{C_p(T,0) - C_p(T,H)}{T}; \quad (21)$$

thus $-\partial \Delta S_M / \partial T > 0$ implies $C_p(T,H) < C_p(T,0)$ for $T < T_{peak}$, and $-\partial S_M / \partial T < 0$ implies $C_p(T,H) > C_p(T,0)$ for $T > T_{peak}$. Taking the formula in Eq. (20) into account this observation affects the sign of $\Delta Q$. The possible regimes are collected in Table 4. Regimes 1 and 2 are straightforward, but regime 3 deserves some comment. In this case quantity $\Delta Q$ may be divided in two components, a negative component $\Delta Q^-$ and a positive component $\Delta Q^+$. It is $\Delta Q^+$ that must be released in full to the cold reservoir, because it cannot be compensated by $\Delta Q^-$ appearing in the separate part of the cycle. Thus the net cooling quantity is equal to $Q_L = Q_C - \Delta Q^+$ in this regime.

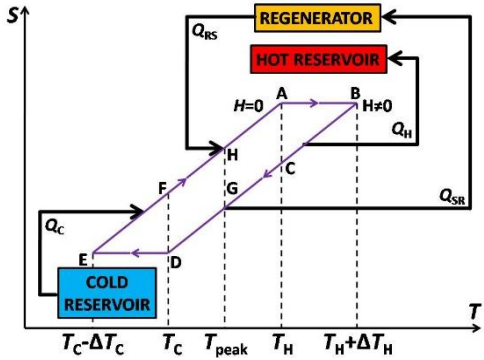 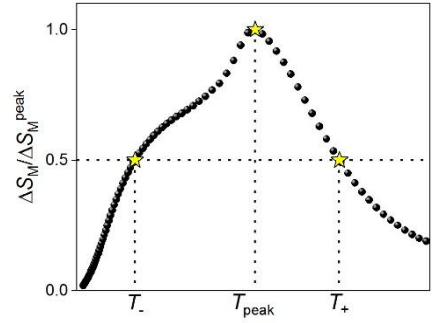

**Fig. 23:** Scheme of the regenerative Brayton cycle.

**Fig. 24:** The range of variation of $T_C$ and $T_H$ is defined by the full width at half maximum of the isothermal entropy change $|\Delta S_M|$ for a given applied field change value.

Apart from the net cooling quantity $Q_L$ we shall discuss the coefficient of performance COP and the cycle efficiency $\eta$. COP is defined as the ratio of $Q_L$ and the work input $W$ for which the first law of thermodynamics implies the following equality $W = -(Q_H + Q_{SR} + Q_C + Q_{RS})$, i.e.

$$\text{COP} = \frac{Q_L}{W}. \tag{22}$$

The cycle efficiency $\eta$ stems from the comparison of its COP with that of the corresponding Carnot cycle working between the same temperatures $T_C$ and $T_H$, i.e.

$$\eta = \frac{\text{COP}}{\text{COP}_{\text{Carnot}}} \cdot 100\%, \text{ where } \text{COP}_{\text{Carnot}} = \frac{T_C}{T_H - T_C}. \tag{23}$$

The range of variation of $T_C$ and $T_H$ is defined by the full width at half maximum of the isothermal entropy change $|\Delta S_M|$ for a given applied field change value (see Fig. 24), i.e. $T_C \in [T_-, T_+ - \Delta T_{\text{off}}]$ and $T_H \in [T_C + \Delta T_{\text{off}}, T_+]$, where the temperature interstice $\Delta T_{\text{off}} = 5$ K. A dense temperature sampling was applied using the data interpolation where necessary.

**Table 4:** The different regimes concerning the sign of the redundant regenerative heat $\Delta Q$.

| Regime No. | Conditions for $T_{C/H}$ | $\Delta Q$ | $Q_L$ |
|---|---|---|---|
| 1 | $T_{C/H} < T_{peak}$ | $\Delta Q < 0$ | $Q_C$ |
| 2 | $T_{C/H} > T_{peak}$ | $\Delta Q > 0$ | $Q_C - \Delta Q$ |
| 3 | $T_C \leq T_{peak} \leq T_H$ | $\Delta Q = \Delta Q^- + \Delta Q^+$ <br> $\Delta Q^- = \int_{T_C}^{T_{peak}} [C_p(T,H) - C_p(T,0)]dT < 0$ <br> $\Delta Q^+ = \int_{T_{peak}}^{T_H} [C_p(T,H) - C_p(T,0)]dT > 0$ | $Q_C - \Delta Q^+$ |

Firstly, let us look at Fig. 25 where the field change dependence of COP and $\eta$ are shown for three most natural choices of $T_C$ and $T_H$, i.e. $T_C = T_-$, $T_H = T_{peak}$ (1), $T_C = T_{peak}$, $T_H = T_+$ (2), $T_C = T_-$, $T_H = T_+$ (3), which correspond to the three regimes in Table 4. It turns out that in regime (3) the Brayton cycle is ineffective being characterized by a negative COP (heating instead of cooling) for all field change values (see the central plot in Fig. 25). Only by confining the working temperatures $T_C$ and $T_H$ of the cycle to either the left (case (1)) or the right (case (2)) wing of the isothermal entropy change peak makes it work properly. In these cases COP is positive and is a decreasing function of the applied field change. For the right wing cycles (case (2)) the COP values are higher than for the left wing cycles (case (1)), however the latter cycles show higher efficiencies with an apparent maximum for intermediate field change values of about 2 – 3 T (cf. the right axes in plots of Fig. 25). In both cases the efficiencies corresponding to the highest field change value of 9 T are the lowest.

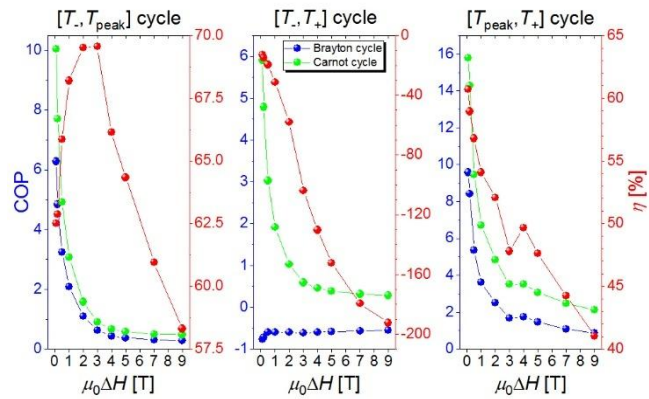

**Fig. 25:** The field change dependence of the Brayton cycle COP and $\eta$ for three most natural choices of $T_C$ and $T_H$ corresponding to the three disparate regimes in Table 4. Left: regime (1); Center: regime (3); Right: regime (2). The green symbols show COP for the corresponding Carnot cycles.

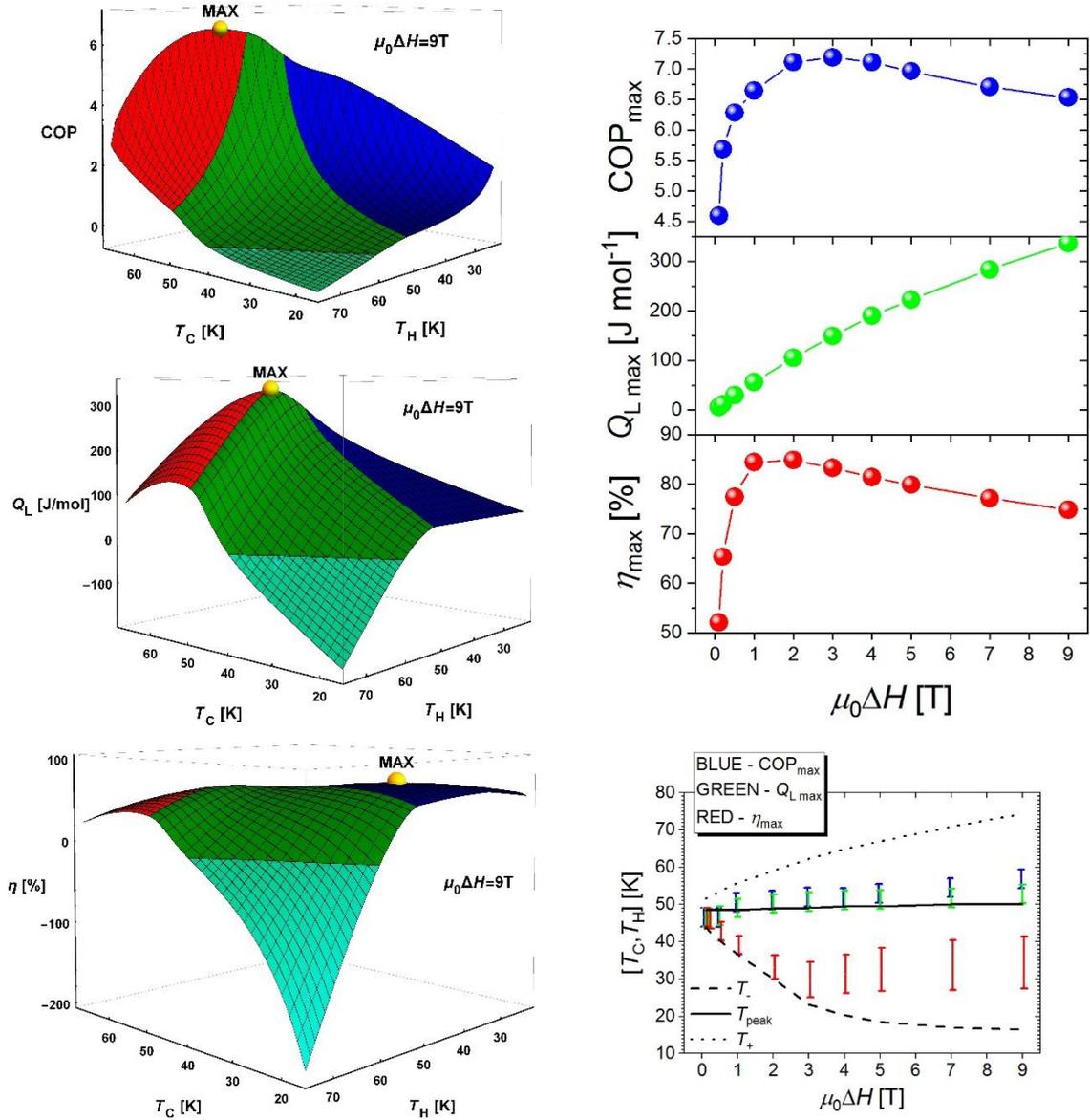

**Fig. 26:** Left: The working temperature map of the Brayton cycle COP (top), $Q_L$ (center), and $\eta$ (bottom) for the maximal field change value of 9 T. The colors correspond to the three $\Delta Q$ regimes; blue – regime (1), red – regime (2), green – regime (3). Right: (top) The applied field change dependence of the maximal values of the Brayton cycle COP, $Q_L$, and $\eta$ found on scanning the working temperatures $(T_C, T_H) \in [T_-, T_+ - \Delta T_{\text{off}}] \times [T_C + \Delta T_{\text{off}}, T_+]$. (bottom) The applied field change dependence of the working temperature intervals $[T_C, T_H]$ for which the maximal values of COP, $Q_L$, and $\eta$ are attained. The solid, dashed and dotted lines corresponding to $T_{\text{peak}}$, $T_-$, and $T_+$, respectively, help to localize the three $\Delta Q$ regimes.

The left column in Fig. 26 shows the result of scanning the working temperature plane $(T_C, T_H) \in [T_-, T_+ - \Delta T_{\text{off}}] \times [T_C + \Delta T_{\text{off}}, T_+]$ for the coefficient of performance COP (top), the net cooling quantity $Q_L$ (center), and the cycle efficiency $\eta$ (bottom) for the largest applied field change value of 9 T. The colored areas of the surface plots correspond to the three disparate

regimes for the redundant regenerative heat $\Delta Q$, i.e. regime (1) – blue, regime (2) – red, and regime (3) – green (cf. Table 4). It can be seen that the largest values of these quantities are assumed at the rear edge of the surfaces which corresponds to the shortest $[T_C, T_H]$ intervals. For longer working temperature intervals, that is while moving to the front corner of the plots, the values of COP, $Q_L$, and $\eta$ gradually decrease to become negative for the longest $[T_C, T_H]$ intervals belonging to regime (3), which was marked by the light green color. The yellow balls show the positions of the global maxima of the three quantities. It is apparent that they are located within different regimes, i.e. in regime (2) for the maximum of COP, at the border of regime (3) for that of $Q_L$, and in regime (1) for that of $\eta$. The behavior of COP, $Q_L$, and $\eta$ as functions of the working temperature pair $(T_C, T_H)$ shown in Fig. 26 is typical, i.e. it is similar for the other field change values. The top right plot in Fig. 26 shows the applied field change dependence of the global maxima of COP, $Q_L$, and $\eta$. It can be seen that $Q_{Lmax}$ is a monotonically increasing function of the field change, while $COP_{max}$ and $\eta_{max}$ attain a maximum value for the intermediate field change values between 1 and 3 T. The bottom right plot shows the intervals $[T_C, T_H]$ corresponding to the global maxima of COP, $Q_L$, and $\eta$ as a function of the applied field change. It can be seen that the corresponding intervals are relatively short. The intervals corresponding to $COP_{max}$ and $Q_{Lmax}$ located in regimes (2) and (3) tend to overlap, while the $\eta_{max}$ intervals, apart from the lowest field change values, are relatively wider and located separately in regime (1). This means that while fixing the working conditions of a Brayton cycle one must decide between a high COP value or a high cycle efficiency. Let us finally note that the shape of the data presented in Fig. 26 depends critically on the value of the $\Delta T_{off}$ parameter defining the narrowest $[T_C, T_H]$ interval considered. Hence the apparent contradiction with the results depicted in Fig. 25 showing higher values of COP for the lowest field change values than those indicated by the top right plot of Fig. 26. This is due to the fact that for the lowest field change values the highest COP values are associated with very narrow $[T_C, T_H]$ intervals resulting in a relatively small denominator in the formula of Eq. (22). The assumed value of $\Delta T_{off} = 5$ K seems to us the most plausible lower limit from the technological point of view.

7.2 The regenerative Ericsson cycle

The scheme of the regenerative Ericsson cycle $A \rightarrow B \rightarrow C \rightarrow D \rightarrow A$ is shown in Fig. 27. It consists of two isothermal processes: $A \rightarrow B$ at $T_H$ and $C \rightarrow D$ at $T_C$, and two isofield processes: $B \rightarrow C$ in nonzero applied field and $D \rightarrow A$ in zero applied field. Like in Section 7.1 $T_C$ and $T_H$ denote the temperatures of the cold and hot reservoir, respectively. The subsequent processes $A \rightarrow B$ and $B \rightarrow C$ involve the following heat quantities:

$$Q_H = \int_{A \rightarrow B} T dS = T_H \Delta S_M(T_H, 0 \rightarrow H) < 0,$$

$$Q_{SR} = \int_{B \rightarrow C} T dS = \int_{T_H}^{T_C} T \frac{\partial S(T, H)}{\partial T} dT = -\int_{T_C}^{T_H} C_p(T, H) dT < 0. \qquad (24)$$

They are both negative, which means that they represent the heat released from the working substance to the hot reservoir and the regenerator, respectively. The following process $C \rightarrow D$ and $D \rightarrow A$ are characterized by the following heat amounts:

$$Q_C = \int_{C \to D} T dS = T_C \Delta S_M(T_C, H \to 0) > 0,$$

$$Q_{RS} = \int_{D \to A} T dS = \int_{T_C}^{T_H} T \frac{\partial S(T,0)}{\partial T} dT = \int_{T_C}^{T_H} C_p(T,0) dT > 0. \quad (25)$$

They are both positive and represent the heat absorbed by the working substance from the cold reservoir and the regenerator, respectively. Similarly to the Brayton cycle, the efficiency of the Ericsson regenerative cycle is affected by the redundant regenerative heat $\Delta Q$ given by the same Eq. (20). Likewise the net cooling quantity $Q_L$ is equal to $Q_C$ corrected by the positive component of $\Delta Q$ and the three disparate regimes listed in Table 4 hold also for the Ericsson regenerative cycles. The coefficient of performance and the cycle efficiency can be calculated using the same Eqs. (22) and (23). The range of variation of $T_C$ and $T_H$ is exactly the same as for the regenerative Brayton cycle discussed in the previous section.

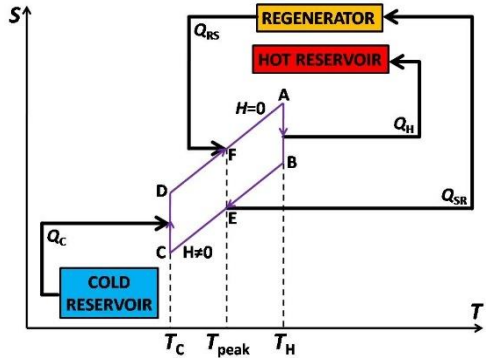
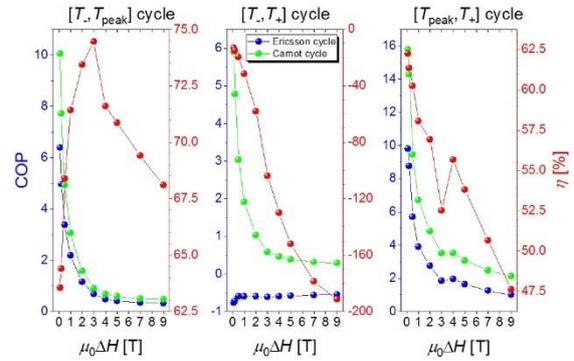

**Fig. 27:** Scheme of the regenerative Ericsson cycle.

**Fig. 28:** The field change dependence of the Ericsson cycle COP and $\eta$ for three most natural choices of $T_C$ and $T_H$ corresponding to the three disparate regimes in Table 4. Left: regime (1); Center: regime (3); Right: regime (2). The green symbols show COP for the corresponding Carnot cycles.

For the sake of comparison exactly the same set of characteristics has been calculated for the Ericsson regenerative cycle. They are shown in Fig. 28 and 29. The behavior of COP and $\eta$ calculated for the three most natural choices of the working temperatures $T_C$ and $T_H$ (cf. Fig. 25 and 28) is qualitatively the same. The Ericsson cycle becomes similarly ineffective for the widest interval $[T_C = T_-, T_H = T_+]$ of regime (3), while COP and $\eta$ are both decreasing functions of the applied field change in the two remaining regimes. The difference is in the magnitudes of these quantities with the efficiencies of the Ericsson cycles exceeding those of the Brayton cycles by ca. 5 % (note the scale ranges of the right ordinate axes of plots in Fig. 25 and 28). The same observation holds for COP and $\eta$ calculated for the highest applied field change of 9 T for different $[T_C, T_H]$ intervals (cf. left top and bottom plots of Fig. 26 and 29) or the global

maxima of COP and $\eta$ shown in the right top plots of Fig. 26 and 29. At the same time the magnitude differences are less apparent for the net cooling quantity $Q_L$ of both cycles (see left center and right top plots of Fig. 26 and 29). Similarly to the Brayton cycles there is an area of $[T_C, T_H]$ intervals contained in regime (3) where the Ericsson cycles become ineffective leading to heating instead of cooling performance (see the light green parts of the surface plots in Fig. 26 and 29). Another important similarity is the fact that the largest values of COP, $Q_L$, and $\eta$ are assumed at the rear edges of the surface plots in Fig. 26 and 29 meaning that it is the short $[T_C, T_H]$ intervals that correspond to the most effective cycles. There are qualitative differences in the field change dependences of COP and $\eta$ corresponding to the global maxima of the working temperature scans (see right top plots of Fig. 26 and 29). While for the Brayton cycle these quantities were maximized for intermediate values of the applied field change, for the Ericsson cycle $COP_{max}$ displays a monotonic increase while $\eta_{max}$ develops a plateau above 3 T. At the same time the field change dependence of $Q_{L\,max}$ looks quite similar for both the cycles. The $[T_C, T_H]$ intervals corresponding to the global maxima of COP, $Q_L$, and $\eta$ are well separated for the higher field change values in the case of the Ericsson cycles locating in regime (2), (3), and (1), respectively (see right bottom plots of Fig. 26 and 29). Moreover, the widths of the working temperature intervals of $\eta_{max}$ are relatively larger for the Brayton cycles. In conclusion, it is the narrow working temperature intervals that make both the cycles most effective with the Ericsson cycle being slightly more efficient than the Brayton cycle.

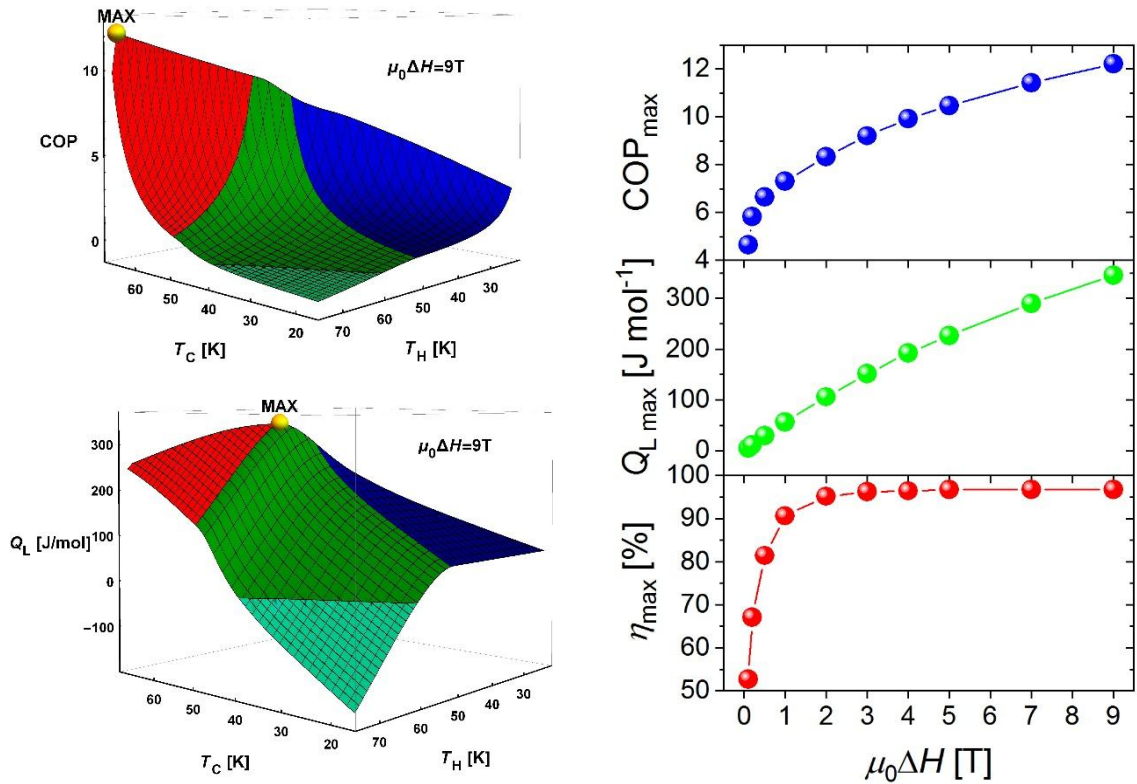

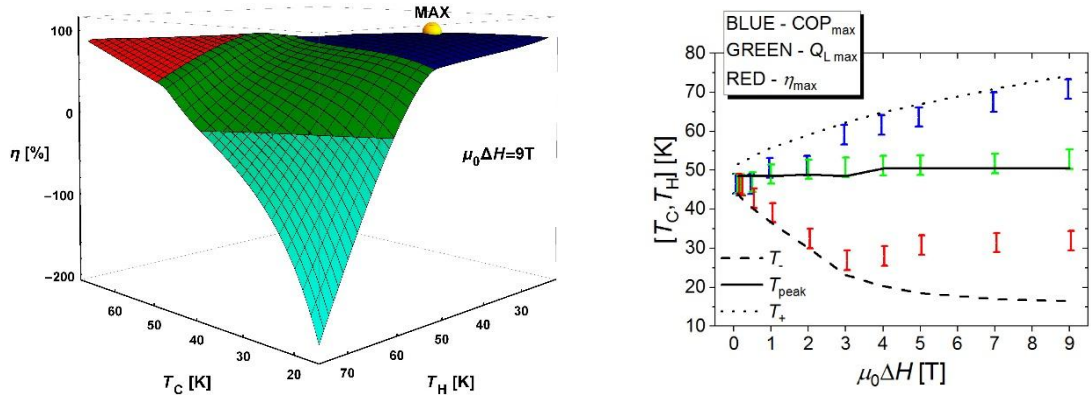

**Fig. 29:** Left: The working temperature map of the Ericsson cycle COP (top), $Q_L$ (center), and $\eta$ (bottom) for the maximal field change value of 9 T. The colors correspond to the three $\Delta Q$ regimes; blue – regime (1), red – regime (2), green – regime (3). Right: (top) The applied field change dependence of the maximal values of the Ericsson cycle COP, $Q_L$, and $\eta$ found on scanning the working temperatures $(T_C, T_H) \in [T_-, T_+ - \Delta T_{\text{off}}] \times [T_C + \Delta T_{\text{off}}, T_+]$. (bottom) The applied field change dependence of the working temperature intervals $[T_C, T_H]$ for which the maximal values of COP, $Q_L$, and $\eta$ are attained. The solid, dashed and dotted lines corresponding to $T_{\text{peak}}$, $T_-$, and $T_+$, respectively, help to localize the three $\Delta Q$ regimes.

## 8. Conclusions

- The comprehensive thermodynamic study of the molecular magnet $[Nb^{IV}\{(\mu\text{-}CN)_4Mn^{II}(H_2O)_2]\}_2 \cdot 4H_2O]_n$ with unblocked connectivity by means of the relaxation calorimetry was reported. Magnetocaloric quantities were determined as a function of the applied field change and temperature. The compound was demonstrated to display typical values of the isothermal entropy change and the adiabatic temperature change for its class of molecular systems.
- The self-consistent scheme based on the magnetic entropy counting was developed to determine the lattice contribution to the total heat capacity (the baseline). It can be applied to other compounds with the second-order phase transition and negligible magnetoelastic coupling, i.e. $\partial C_p(\text{lattice})/\partial H \ll 1$. This approach enables one to determine the total entropy within the full range of temperatures starting at 0 K up to the maximal detected temperature and hence to consistently calculate the magnetocaloric properties.
- The fact that temperatures at which $\Delta S_M$ and $\Delta T_{ad}$ have peaks are shifted off the transition temperature $T_c$ towards higher temperatures was shown to be consistent with the theoretical predictions based on the definitions of these quantities.
- The molecular field approximation was employed to calculate the temperature and field dependence of the isothermal magnetic entropy change for the spin network corresponding to the studied system with and without magnetic interactions. The differences with the experimental output were discussed in terms of the spin fluctuations absent in the mean-field model. The purely paramagnetic counterpart of the studied compound was shown to considerably exceed the magnitude of its entropic effect with

- the corresponding peak localized at 0 K, which might not always be of practical importance.
- For the lowest applied field change values the inverse magnetocaloric effect was revealed in the lowest temperature regime. This feature seems to be characteristic of systems with antiferromagnetic coupling.
- The field dependence of the isothermal magnetic entropy change $\Delta S_\mathrm{M}$ was analyzed revealing that at no temperature the value of exponent $n$ is universal, i.e. field independent. This makes the confirmation of the theoretical prediction for the value of $n$ at $T_\mathrm{c}$ or the indication of a valid universality class for the critical behavior highly ambiguous.
- The effectiveness of the Brayton and Ericsson thermodynamic cycles with the studied compound employed as the working substance was compared and discussed. The major conclusion is that it is narrow temperature ranges of operation that ensure the optimal efficiencies, which suggests to use a serial connection of refrigeration units (a cascade system) with each unit having its own optimally adapted working temperatures filling subsequently the target temperature range.

**References**


[1] O. Kahn, Molecular Magnetism, VCH Publishers, Inc.: Weinheim, Germany, 1993.

[2] B. Sieklucka, D. Pinkowicz, Molecular Magnetic Materials: Concepts and Applications, Wiley-VCH Verlag GmbH & Co. KGaA, Weinheim, Germany, 2017.

[3] D. Pinkowicz, B. Czarnecki, M. Reczyński, M. Arczyński, Multifunctionality in molecular magnets, Sci. Progress **98** (2015) 346-378, DOI: 10.3184/003685015X14465681600416

[4] J. Ferrando-Soria, J. Vallejo, M. Castellano, J. Martínez-Lillo, E. Pardo, J. Cano, I. Castro, F. Lloret, R. Ruiz-García, M. Julve, Molecular magnetism *quo vadis?* A historical perspective from a coordination chemist viewpoint, Coord. Chem. Rev. **339** (2017) 17-103, DOI: 10.1016/j.ccr.2017.03.004

[5] F.-S. Guo, A. K. Bar, R. A. Layfield, Main group chemistry at the interface with molecular magnetism, Chem. Rev. **119** (2019) 8479-8505, DOI: 10.1021/acs.chemrev.9b00103

[6] E. Coronado, Molecular magnetism: from chemical design to spin control in molecules, materials and devices, Nat. Rev. Mater. **5** (2020) 87-104, DOI: 10.1038/s41578-019-0146-8

[7] D. Gatteschi, Molecular magnetism: a basis for new materials, Adv. Mater. **6** (1994) 635–645, DOI: 10.1002/adma.19940060903

[8] R. Sessoli, D. Gatteschi, A. Caneschi, M. A. Novak, Magnetic bistability in a metal-ion cluster, Nature **365** (1993) 141–143, DOI: 10.1038/365141a0

[9] L. Thomas, Macroscopic quantum tunnelling of magnetization in a single crystal of nanomagnets, Nature **383** (1996) 145–147, DOI: 10.1038/383145a0

[10] O. Sato, T. Iyoda, A. Fujishima, K. Hashimoto, Photoinduced magnetization of a cobalt–iron cyanide, Science **272** (1996) 704–705, DOI: 10.1126/science.272.5262.70



[11] E. Coronado, M. C. Giménez-López, G. Levchenko, F. M. Romero, V. Garcia-Baonza, A. Milner, M. Paz-Pasternak, Pressure- tuning of magnetism and linkage isomerism in iron(II) hexacyanochromate, J. Am. Chem. Soc. **127** (2005) 4580–4581, DOI: 10.1021/ja043166z

[12] S. Ohkoshi, S. Takano, K. Imoto, M. Yoshikiyo, A. Namai, H. Tokoro, 90-Degree optical switching of output second-harmonic light in chiral photomagnet, Nat. Photonics **8** (2014) 65–71, DOI: 10.1038/nphoton.2013.310

[13] G. Abellán, C. Martí-Gastaldo, A. Ribera, E. Coronado, Hybrid materials based on magnetic layered double hydroxides: a molecular perspective, Acc. Chem. Res. **48** (2015) 1601–1611, DOI: 10.1021/acs.accounts.5b00033

[14] M. Liberka, M. Zychowicz, S. Chorąży, Solvato- and vapochromic exchange coupled $Dy_2$ single molecule magnets achieved by attaching iron-cyanido metalloligands, Inorg. Chem. Frontiers **11** (2024) 2081-2097, DOI: 10.1039/d4qi00138a

[15] J. J. Zakrzewski, M. Liberka, J. Wang, S. Chorąży, S. Ohkoshi, Optical phenomena i molecule-based magnetic materials, Chem. Rev. **124** (2024) 5930-6050, DOI: 10.1021/acs.chemrev.3c00840

[16] M. Clemente-León, E. Coronado, M. López-Jordà, C. Desplanches, S. Asthana, H. Wang, J.-F. Létard, A hybrid magnet with coexistence of ferromagnetism and photo-induced Fe(III) spin-crossover, Chem. Sci. **2** (2011) 1121-1127, DOI: 10.1039/c1sc00015b

[17] S. Chorąży, J. J. Zakrzewski, M. Magott, T. Korzeniak, B. Nowicka, D. Pinkowicz, R. Podgajny, B. Sieklucka, Octacyanidometallates for multifunctional molecule-based materials, Chem. Soc. Rev. **49** (2020) 5945-6001, DOI: 10.1039/d0cs00067a

[18] B. Nowicka, T. Korzeniak, O. Stefańczyk, D. Pinkowicz, S. Chorąży, R. Podgajny, B. Sieklucka, The impact of ligands upon topology and functionality of octacyanidometallate-based assemblies, Coord. Chem. Rev. **256** (2012) 1946-1971, DOI: 10.1016/j.ccr.2012.04.008

[19] M. Reczyński, D. Pinkowicz, K. Nakbayashi, C. Näther, J. Stanek, M. Kozieł, J. Kalinowska-Tłuścik, B. Sieklucka, S. Ohkoshi, B. Nowicka, Room-temperature bistability in a Ni-Fe chain: Electron transfer controlled by temperature, pressure, light and humidity, Angew. Chem. Int. Ed. **60** (2021) 2330-2338, DOI: 10.1002/anie.202012876

[20] B. Sieklucka, R. Podgajny, T. Korzeniak, B. Nowicka, D. Pinkowicz, M. Kozieł, A decade of octacyanides in polynuclear molecular materials, Eur. J. Inorg. Chem. **2011**, 305-326, DOI: 10.1002/ejic.201001055

[21] D. Maspoch, D. Ruiz-Molina, K. Wurst, N. Domingo, M. Cavallini, F. Biscarini, J. Tejada, C. Rovira, J. Veciana, A nanoporous molecular magnet with reversible solvent-induced mechanical and magnetic properties, Nat. Mater. **2** (2003) 190–195, DOI: 10.1038/nmat834

[22] B. Nowicka, M. Heczko, M. Reczyński, M. Rams, B. Gaweł, W. Nitek, B. Sieklucka, Exploration of a new building block for the construction of cyano-bridged solvatomagnetic assemblies: $[Ni(cyclam)]^{3+}$, CrystEngComm. **18** (2016) 7011-7020, DOI: 10.1039/c6ce01524g

[23] O. Stefańczyk, S. Ohkoshi, Humidity – a powerful tool to customize the physical properties of molecular magnets, Chem.-A Eur. J. **25** (2019) 15963-15977, DOI: 10.1002/chem.201903586



[24] W. Kosaka, Z. Liu, J. Zhang, Y. Sato, A. Hori, R. Matsuda, S. Kitagawa, H. Miyasaka, Gas-responsive porous magnet distinguishes the electron spin of molecular oxygen, Nat. Commun. **9** (2018) 5420, DOI: 10.1038/s41467-018-07889-1

[25] E. Coronado, J. R. Galán-Mascarós, C. J. Gómez-García, V. Laukhin, Coexistence of ferromagnetism and metallic conductivity in a molecule-based layered compound, Nature **408** (2000) 447–449, DOI: 10.1038/35044035

[26] M. Reczyński, M. Heczko, M. Kozieł, S. Ohkoshi, B. Sieklucka, B. Nowicka, Proton-conducting humidity-sensitive $Ni^{II}$-$Nb^{IV}$ magnetic coordination network, Inorg. Chem. **58** (2019) 15812-15823, DOI: 10.1021/acs.inorgchem.9b02141

[27] Z. Wang, Y. Huang, W. Gong, Q. Yan, S. Ren, Lithiation bridged molecular conducting magnets, Applied Materials Today **38** (2024) 102188, DOI: 10.1016/j.apmt.2024.102188

[28] E. Coronado, C. J. Gómez-García, A. Nuez, F. M. Romero, J. C. Waerenborgh, Synthesis, chirality and magnetic properties of bimetallic cyanide-bridged two-dimensional ferromagnets, Chem. Mater. **18** (2006) 2670–2681, DOI: 10.1021/cm0600879

[29] C. Train, R. Gheorghe, V. Krstic, L.-M. Chamoreau, N. S. Ovanesyan, Geert L. J. A. Rikken, M. Gruselle, M. Verdaguer, Strong magneto-chiral dichroism in enantiopure chiral ferromangets, Nat. Mater. **7** (2008) 729–734, DOI: 10.1038/nmat2256

[30] K. Inoue, Chiral magnetism: coupling static and dynamic chirality, Chem. Lett. **50** (2021) 742-751, DOI: 10.1246/cl.200840

[31] F. Houard, Q. Evrard, G. Calvez, Y. Suffren, C. Daiguebonne, O. Guillou, F. Gendron, B. Le Guennic, T. Guizouarn, V. Dorcet, M. Mannini, K. Bernot, Chiral supramolecular nanotubes of single-chain magnets, Angew. Chem. Int. Ed. **59** (2020) 780-784, DOI: 10.1002/anie.201913019

[32] G. Handzlik, K. Rzepka, D. Pinkowicz, The underexplored field of lanthanide complexes with helicene ligands: Towards chiral lanthanide single molecule magnets, Magnetochemistry **7** (2021) 138, DOI: 10.3390/magnetochemistry7100138

[33] M. Zeng, L. Miao, X.-R. Wu, C.-M. Liu, H.-Z. Kou, Chiral Dy(III) fluorescent single-molecule magnet based on an achiral flexible ligand, Magnetochemistry **8** (2022) 166, DOI: 10.3390/magnetochemistry8120166

[34] M. Atzori, Geert L. J. A. Rikken, C. Train, Magneto-chiral dichroism: a playground for molecular chemists, Chem.-A Eur. J. **26** (2020) 9784-9791, DOI: 10.1002/chem.202000937

[35] M. Atzori, I. Breslavetz, K. Paillot, Geert L. J. A. Rikken, C. Train, Role of structural dimensionality in the magneto-chiral dichroism of chiral molecular ferrimagnets, J. Mater. Chem. C **10** (2022) 13939-13945, DOI: 10.1039/d2tc01777f

[36] M. Kurmoo, A. W. Graham, P. Day, S. J. Coles, M. B. Hursthouse, J. L. Caulfield, J. Singleton, F. L. Pratt, W. Hayes, L. Ducasse, P. Guionneau, Superconducting and semiconducting magnetic charge transfer salts: $(BEDT-TTF)_4AFe(C_2O_4)_3 \cdot C_6H_5CN$ (A = $H_2O$, K, $NH_4$), J. Am. Chem. Soc. **117** (1995) 12209–12217, DOI: 10.1021/ja00154a022



[37] S. Fukuoka, S. Fukuchi, H. Akutsu, A. Kawamoto, Y. Nakazawa, Magnetic and electronic properties of π-*d* interacting molecular magnetic superconductor κ-(BETS)$_2$FeX$_4$ (X=Cl, Br) studied by angle-resolved heat capacity measurements, Crystals **9** (2019) 66, DOI: 10.3390/cryst9020066

[38] H.-N. Xia, E. Minamitani, R. Žitko, Z.-Y. Liu, X. Liao, M. Cai, Z.-H. Ling, W.-H. Zhang, S. Klyatskaya, M. Ruben, Y.-S. Fu, Spin-orbital Yu-Shiba-Rusinov states in single Kondo molecular magnet, Nat. Commun. **13** (2022) 6388, DOI: 10.1038/s41467-022-34187-8

[39] E. Moreno-Pineda, W. Wernsdorfer, Measuring molecular magnets for quantum technologies, Nat. Rev. Phys. **3** (2021) 645-659, DOI: 10.1038/s42254-021-00340-3

[40] E. Moreno-Pineda, W. Wernsdorfer, Magnetic Molecules as building blocks for quantum technologies, Adv. Quantum Technol. **2024**, 2300367, DOI: 10.1002/qute.202300367

[41] G. Gabarró-Riera, E. C. Sañudo, Challenges for exploiting nanomagnet properties on surfaces, Commun. Chem. **7** (2024) 99, DOI: 10.1038/s42004-024-01183-6

[42] B. R. Dahal, M. Savadkoohi, A. Grizzle, C. D'Angelo, V. Lamberti, P. Tyagi, Easy axis anisotropy creating high contrast magnetic zones on magnetic tunnel junctions based molecular spintronics devise (MTJMSD), Sci. Rep. **12** (2022) 5721, DOI: 10.1038/s41598-022-09321-7

[43] E. Mutunga, C. D'Angelo, P. Tyagi, Magnetic molecules lose identity when connected to different combinations of magnetic metal electrodes in MTJ-based molecular spintronics devices (MTJMSD), Sci. Rep. **13** (2023) 16201, DOI: 10.1038/s41598-023-42731-9

[44] J. Liu, J. Mrozek, A Ullah, Y. Duan, J. J. Baldoví, E. Coronado, A. Gaita-Ariño, A. Ardavan, Quantum coherent spin-electric control in a molecular nanomagnet at clock transitions, Nat. Phys. **17** (2021) 1205-1209, DOI: 10.1038/s41567-021-01355-4

[45] G. M. Gutiérrez-Finol, S. Giménez-Santamarina, Z. Hu, L. E. Rosaleny, S. Cardona-Serra, A. Gaita-Ariño, Lanthanide molecular nanomagnets as probabilistic bits, NPJ Comp. Mater. **9** (2023) 196, DOI: 10.1038/s41524-023-01149-7

[46] CY. Yang, ZX. Chen, CJ. Yu, JW. Cao, GJ. Ke, WY. Zhu, WX. Liang, JX. Huang, WQ. Cai, C. Saha, M. A. Sabuj, N. Rai, XX. Li, JL. Yang, Y. Li, F. Huang, XF. Guo, Regulation of quantum spin conversions in a single molecular radical, Nat. Nanotech. **19** (2024) 978-985, DOI: 10.1038/s41565-024-01632-2

[47] V. E. Campbell, M. Tonelli, I. Cimatti, J.-B. Moussy, L. Tortech, Y. J. Dappe, E. Rivière, R. Guillot, S. Delprat, R. Mattana, P. Seneor, P. Ohresser, F. Choueikani, E. Otero, F. Koprowiak, V. G. Chilkuri, N. Suaud, N. Guihéry, A. Galtayries, F. Miserque, M.-A. Arrio, P. Sainctavit, T. Mallah, Engineering the magnetic coupling and anisotropy at the molecule-magnetic surface interface in molecular spintronic devices, Nat. Commun. **7** (2016) 13646, DOI: 10.1038/ncomms13646

[48] C. Li, C. Kaspar, P. Zhou, J.-C. Liu, O. Chahib, T. Glatzel, R. Haener, U. Aschauer, S. Decurtins, S.-X. Liu, M. Thoss, E. Meyer, R. Pawlak, Strong signature of electron-vibration coupling in molecules on Ag(111) triggered by tip-gated discharging, Nat. Commun. **14** (2023) 5956, DOI: 10.1038/s41467-023-41601-2



[49] A. Candini, D. Klar, S. Marocchi, V. Corradini, R. Biagi, V. De Renzi, U. del Pennino, F. Troiani, V. Bellini, S. Klyatskaya, M. Ruben, K. Kummer, N. B. Brookes, H. Huang, A. Soncini, H. Wende, M. Affronte, Spin-communication channels between Ln(III) bis-phthalocyanines molecular nanomagnets and a magnetic substrate, Sci, Rep. **6** (2016) 21740, DOI: 10.1038/srep21740

[50] T. Gadzikwa, Interweaving different metal-organic frameworks, Nat. Chem. **15** (2023) 1324-1326, DOI: 10.1038/s41557-023-01335-6

[51] GZ. Liu, YA. Guo, CL. Chen, Y. Lu, GN. Chen, GP. Liu, Y. Han, WQ. Jin, NP. Xu, Eliminating lattice defects in metal-organic framework molecular sieving membranes, Nat. Mat. **22** (2023) 769-776, DOI: 10.1038/s41563-023-01541-0

[52] J. Zhang, W. Kosaka, Y. Kitagawa, H. Miyasaka, A metal-organic framework that exhibits $CO_2$-induced transitions between paramagnetism and ferrimagnetism, Nat. Chem. **13** (2021) 191-199, DOI: 10.1038/s41557-020-00577-y

[53] B. Field, A. Schiffrin, N. V. Medhekar, Correlation-induced magnetism in substrate-supported 2D metal-organic frameworks, NPJ Comp. Mater. **8** (2022) 227, DOI: 10.1038/s41524-022-00918-0

[54] S. S. Rajasree, XL. Li, P. Deria, Physical properties of porphyrin-based crystalline metal-organic framework, Commun. Chem. **4** (2021) 47, DOI: 10.1038/s42004-021-00484-4

[55] DG. Ha, RM. Wan, CA. Kim, TA. Lin, LM. Yang, T. Van Voorhis, M. A. Baldo, M. Dinca, Exchange controlled triplet fusion in metal-organic frameworks, Nat. Mat. **21** (2022) 1275-1281, DOI: 10.1038/s41563-022-01368-1

[56] P. Matvija, F. Rozboril, P. Sobotík, I. Ost'ádal, B. Pieczyrak, L. Jurczyszyn, P. Kocán, Electric-field-controlled phase transition in a 2D molecular layer, Sci. Rep. **7** (2017) 7357, DOI: 10.1038/s41598-017-07277-7

[57] R. Torres-Cavanillas, A. Forment-Aliaga, Design of stimuli-responsive transition metal dichalcogenides, Commun. Chem. **7** (2024) 241, DOI: 10.1038/s42004-024-01322-z

[58] M. Magott, M. Brzozowska, S. Baran, V. Vieru, D. Pinkowicz, An intermetallic molecular nanomagnet with the lanthanide coordinated only by transition metals, Nat. Commun. **13** (2022) 2014, DOI: 10.1038/s41467-022-29624-7

[59] D. Errulat, K. L. M. Harriman, D. A. Galico, E. V. Salerno, J. van Tol, A. Mansikkamäki, M. Rouzière, S. Hill, R. Clérac, M. Murugesu, Slow magnetic relaxation in a europium(II) complex, Nat. Commun. **15** (2024) 3010, DOI: 10.1038/s41467-024-46196-w

[60] L. Zhao, Y.-S. Meng, Q. Liu, O. Sato, Q. Shi, H. Oshio, T. Liu, Switching the magnetic hysteresis of an [$Fe^{II}$-NC-$W^V$]-based coordination polymer by photoinduced reversible spin crossover, Nat. Chem. **13** (2021) 698, DOI: 10.1038/s41557-021-00695-1



[61] Y. Duan, L. E. Rosaleny, J. T. Coutinho, S. Giménez-Santamarina, A. Scheie, J. J. Baldovi, S. Cardona-Serra, A. Gaita-Ariño, Data-driven design of molecular nanomagnets, Nat. Commun. **13** (2022) 7626, DOI: 10.1038/s41467-022-35336-9

[62] ZB. Hu, XY. Yang, JL. Zhang, LA. Gui, YF. Zhang, XD. Liu, ZH. Zhou, YC. Jiang, Y. Zhang, S. Dong, Y. Song, Molecular ferroelectric with low-magnetic-field magnetoelectricity at room temperature, Nat. Commun. **15** (2024) 4702, DOI: 10.1038/s41467-024-49053-y

[63] K. A. Gschneidner Jr., V. K. Pecharsky, The influence of magnetic field on the thermal properties of solids, Materials Science and Engineering **A287** (2000) 301-310, DOI: 10.1016/S0921-5093(00)00788-7

[64] A. M. Tishin, Y. I. Spichkin, The magnetocaloric effect and its applications, IOP Publishing Ltd: London, UK, 2003.

[65] A. Greco, A. R. Farina, C. Masseli, Caloric solid-state magnetocaloric cooling: Physical phenomenon, thermodynamic cycles and materials, Tecnica Italiana - Ital. J. Engineer. Sci. **65** (2021) 58-66, DOI: 10.18280/ti-ijes.650109

[66] A. Alahmer, M. Al.-Amayreh, A. O. Mostafa, M. Al-Dabbas, H. Rezk, Magnetic refrigeration design technologies: state of the art and general perspectives, Energies **14** (2021) 4662, DOI: 10.3390/en14154662

[67] V. K. Pecharsky, K. A. Gschneidner Jr., A. O. Pecharsky, A. M. Tishin, Thermodynamics of the magnetocaloric effect, Phys. Rev. B **64** (2001) 144406, DOI: 10.1103/PhysRevB.64.144406

[68] A. Giguère, M. Foldeaki, B. R. Gopal, R. Chahine, T. K. Bose, A. Frydman, J. A. Barclay, Direct measurement of the "giant" adiabatic temperature change $Gd_5Si_2Ge_2$, Phys. Rev. Lett. **83** (1999) 2262-2265, DOI: 10.1103/PhysRevLett.83.2262

[69] J. Kamarád, J. Kastil, Z. Arnold, Practical system for the direct measurement of magnetocaloric effect by micro-thermocouples, Rev. Sci. Instrum. **83** (2012) 083902, DOI: 10.1063/1.4739962

[70] M. Ghahremani, Y. Jin, L. H. Bennett, E. Della Torre, H. ElBidweihy, S. Gu, Design and instrumentation of an advanced magnetocaloric direct temperature measurement system, IEEE Trans. Magnet. **48** (2012) 3999-4002, DOI: 10.1109/TMAG.2012.2203108

[71] Y. Hirayama, R. Iguchi, X.-F. Miao, K. Hono, K. Uchida, High-throughput direct measurement of magnetocaloric effect based on lock-in thermography technique, Appl. Phys. Lett. **111** (2017) 163901, DOI: 10.1063/1.5000970

[72] S. Ghorai, D. Hedlund, M. Kapuscinski, P. Svedlindh, A setup for direct measurement of the adiabatic temperature change in magnetocaloric materials, IEEE Trans. Instrum. Measur. **72** (2023) 6005109, DOI: 10.1109/TIM.2023.3272387

[73] Y. Zhang, T. Nomoto, S. Yamashita, H. Akutsu, N. Yoshinari, T. Konno, Y. Nakazawa, Direct measurement of magnetocaloric effect (MCE) in frustrated Gd-based molecular complexes, J. Therm. Anal. Calorim. (2024), DOI: 10.1007/s10973-024-13344-9



[74] M. Falsaperna, P. J. Saines, Development of magnetocaloric coordination polymers for low temperature cooling, Dalton Trans. **51** (2022) 3394-3410, DOI: 10.1039/d1dt04073a

[75] A. Dey, P. Bag, P. Kalita, V. Chandrasekhar, Heterometallic $Cu^{II}$-$Ln^{III}$ complexes: Single-molecule magnets and magnetic refrigerants, Coord. Chem. Rev. **432** (2021) 213707, DOI: 10.1016/j.ccr.2020.213707

[76] J.-J. Hu, J. Peng, S.-J. Liu, H.-R. Wen, Recent advances in lanthanide coordination polymers and clusters with magnetocaloric effect or single-molecule magnet behavior, Dalton Trans. **50** (2021) 15473-15487, DOI: 10.1039/d1dt02797b

[77] P. Konieczny, W. Sas, D. Czernia, A. Pacanowska, M. Fitta, R. Pełka, Magnetic cooling: a molecular perspective, Dalton Trans. **51** (2022) 12762-12780, DOI: 10.1039/d2dt01565j

[78] P. Richardson, D. I. Alexandropoulos, L. Cunha-Silva, G. Lorusso, M. Evangelisti, J. Tang, T. C. Stamatatos, 'All three-in-one': ferromagnetic interactions, single-molecule magnetism, and magnetocaloric properties in a new family of [$Cu_4Ln$] ($Ln^{III}$ = Gd, Tb, Dy) clusters, Inorg. Chem. Front. **2** (2015) 945-948, DOI: 10.1039/c5qi00146c

[79] M. Evangelisti, A. Candini, M. Affronte, E. Pasca, L. J. de Jongh, R. T. W. Scott, E. K. Brechin, Magnetocaloric effect in spin-degenerated molecular nanomagnets, Phys. Rev. B **79** (2009) 104414, DOI: 10.1103/PhysRevB.79.104414

[80] E. Manuel, M. Evangelisti, M. Affronte, M. Okubo, C. Train, M. Verdaguer, Magnetocaloric effect in hexacyanochromate Prussian blue analogs, Phys. Rev. B **73** (2006) 172406, DOI: 10.1103/PhysRevB.73.172406

[81] M. Affronte, A. Ghirri, S. Carretta, G. Amoretti, S. Piligkos, G. A. Timco, R. E. P. Winpenny, Engineering molecular rings for magnetocaloric effect, Appl. Phys. Lett. **84** (2004) 3468-3470, DOI: 10.1063/1.1737468

[82] M. Evangelisti, A, Candini, A. Ghirri, M. Affronte, E. K. Brechin, E. J. L. McInnes, Spin-enhanced magnetocaloric effect in molecular nanomagnets, Appl. Phys. Lett. **87** (2005) 072504, DOI: 10.1063/1.2010604

[83] F. Torres, J. M. Hernández, X. Bohigas, J. Tejada, Giant and time-dependent magnetocaloric effect in high-spin molecular magnets, Appl. Phys. Lett. **77** (2000) 3248-3250, DOI: 10.1063/1.1325393

[84] F. Torres, X. Bohigas, J. M. Hernández, J. Tejada, Magnetocaloric effect in $Mn_{12}$ 2-Cl benzoate, J. Phys.: Condens. Matter **15** (2003) L119, DOI: 10.1088/0953-8984/15/4/102

[85] X.-X. Zhang, H.-L. Wei, Z.-Q. Zhang, L.-Y. Zhang, Anisotropic magnetocaloric effect in nanostructured magnetic clusters, Phys. Rev. Lett. **87** (2001) 157203, DOI: 10.1103/PhysRevLett.87.157203

[86] P. Bhatt, N. Maiti, M. D. Mukadam, S. S. Meena, A. Kumar, S. M. Yusuf, Cluster spin-glass state with magnetocaloric effect in open framework structure of the Prussian blue analog molecular magnet $K_{2x/3}Cu[Fe(CN)_6]_{2/3}n H_2O$, Phys. Rev. B **108** (2023) 014412, DOI: 10.1103/PhysRevB.108.014412



[87] R. Pełka, K. Szałowski, M. Rajnák, W. Sas, D. Czernia, P. Konieczny, J. Kobylarczyk, M. Mihálik, P. Kögerler, Low-temperature magnetocaloric effect of the polyoxovanadate molecular magnet {$V^{IV/V}_{12}As_8$}: An experimental study, J. Magn. Magn. Mater. **591** (2024) 171722, DOI: 10.1016/j.jmmm.2024.171722

[88] Q. Zhang, S.-Y. Yang, S.-J. Chen, L. Shi, J. Yang, Z.-F. Tian, D. Shao, Lanthanide dimers in coordination chains constructed by sole dicarboxylate ligand for single-molecule magnet behavior and magnetocaloric effect, J. Mol. Struct. **1294** (2023) 136349, DOI: 10.1016/j.molstruc.2023.136349

[89] G. Lorusso, O. Roubeau, M. Evangelisti, Rotating magnetocaloric effect in an anisotropic molecular dimer, Agew. Chem.-Int. Ed. **55** (2016) 3360-3363, DOI: 10.1002/anie.201510468

[90] P. Konieczny, D. Czernia, T. Kajiwara, Rotating magnetocaloric effect in highly anisotropic $Tb^{III}$ and $Dy^{III}$ single molecular magnets, Sci. Rep. **12** (2022) 16601, DOI: 10.1038/s41598-022-20893-2

[91] P. Danylchenko, R. Tarasenko, E. Čižmár, V. Tkáč, A. Feher, A. Orendáčová, M. Orendáč, Giant rotational magnetocaloric effect in Ni(*en*)(H$_2$O)$_4$·2H$_2$O: Experiment and theory, Magnetochemistry **8** (2022) 39, DOI: 10.3390/magnetochemistry8040039

[92] Y.-X. Shang, Y.-D. Cao, Y.-F. Xie, S.-W. Zhang, P. Cheng, A 1D Mn-based coordination polymer with significant magnetocaloric effect, Polyhedron **202** (2021) 115173, DOI: 10.1016/j.poly.2021.115173

[93] L. Tacconi, A. S. Manvell, M. Briganti, D. Czernia, H. Weihe, P. Konieczny, J. Bendix, M. Perfetti, Exploiting high-order magnetic anisotropy for advanced magnetocaloric refrigerants, Angew. Chem.-Int. Ed. (2024) e202417582, DOI: 10.1002/anie.202417582

[94] M. Bałanda, R. Pełka, M. Fitta, Ł. Laskowski, M. Laskowska, Relaxation and magnetocaloric effect in the Mn$_{12}$ molecular nanomagnet incorporated into mesoporous silica: a comparative study, RSC Adv. **6** (2016) 49179-49186, DOI: 10.1039/c6ra04063b

[95] S. M. Yusuf, A. Kumar, J. V. Jakhmi, Temperature- and magnetic-field-controlled magnetic pole reversal in a molecular magnetic compound, Appl. Phys. Lett. **95** (2009) 182506, DOI: 10.1063/1.3259652

[96] N. Qiao, X.-X. Li, Y. Chen, X.-Y. Xin, C. Yang, S.-S. Dong, Y.-Z. Wang, X.-J. Li, Y.-P. Hua, W.-M. Wang, Three Ln$_2$ compounds (Gd$_2$, Tb$_2$ and Dy$_2$) with a Ln$_2$O$_2$ center showing magnetic refrigeration property and single-molecular magnet behavior, Polyhedron **215** (2022) 115675, DOI: 10.1016/j.poly.2022.115675

[97] L. Regeciová, P. Farkašovský, Quantum design of magnetic structures with enhanced magnetocaloric properties, J. Phys. D.: Appl. Phys. **57** (2024) 455301, DOI: 10.1088/1361-6463/ad5e8f

[98] P. Kowalewska, K. Szałowski, Magnetocaloric properties of V6 molecular magnet, J. Magn. Magn. Mater. **496** (2020) 165933, DOI: 10.1016/j.jmmm.2019.165933

[99] J. M. Florez, A. S. Núñez, C. García, P. Vargas, Magnetocaloric features of complex molecular magnets: The (Cr$_7$Ni)$_2$Cu molecular magnet and beyond, J. Magn. Magn. Mater. **322** (2010) 2810-2818, DOI: 10.1016/j.jmmm.2010.04.035



[100] K. Szałowski, Low-temperature magnetocaloric properties of V12 polyoxovanadate molecular magnet: A theoretical study, Materials **13** (2020) 4399, DOI: 10.3390/ma13194399

[101] K. Szałowski, P. Kowalewska, Magnetocaloric effect Cu5-NIPA molecular magnet: A theoretical study, Materials **13** (2020) 485, DOI: 10.3390/ma13020485

[102] C. Beckmann, J. Ehrens, J. Schnack, Rotational magnetocaloric effect of anisotropic giant-spin molecular magnets, J. Magn. Magn. Mater. **482** (2019) 113-119, DOI: 10.1016/j.jmmm.2019.02.064

[103] B Boughazi, M. Kerouad, Monte Carlo study of magnetic properties and magnetocaloric effect in a molecular-based magnet $AFe^{II}Fe^{III}(C_2O_4)_3$ nanographene, Eur. Phys. J. Plus **136** (2021) 926, DOI: 10.1140/epjp/s13360-021-01916-9

[104] M. Fitta, R. Pełka, M. Bałanda, M. Czapla, M. Mihalik, D. Pinkowicz, B. Sieklucka, T. Wasiutyński, M. Zentková, Magnetocaloric effect in $Mn_2$-pyridazine-$[Nb(CN)_8]$ molecular magnetic sponge, Eur. J. Inorg. Chem. **2012** (2012) 3830-3834, DOI: 10.1002/ejic.201200374

[105] M. Fitta, M. Bałanda, M. Mihalik, R. Pełka, D. Pinkowicz, B. Sieklucka, M. Zentková, Magnetocaloric effect in M-pyrazole-$[Nb(CN)_8]$ (M=Ni, Mn) molecular compounds, J. Phys.: Condens. Matter **24** (2012) 506002, DOI: 10.1088/0953-8984/24/50/506002

[106] M. Fitta, M. Bałanda, R. Pełka, P. Konieczny, D. Pinkowicz, B. Sieklucka, Magnetocaloric effect and critical behavior in $Mn_2$-pyridazine-$[Nb(CN)_8]$ molecular compound under pressure, J. Phys.: Condens. Matter **25** (2013) 496012, DOI: 10.1088/0953-8984/25/49/496012

[107] R. Pełka, Calorimetric studies of octacyanometallate-based coordination polymers displaying long-range magnetic order, Current Inorg. Chem. **4** (2014) 146-166, DOI: 10.2174/1877944104666140825194745

[108] R. Pełka, P. Konieczny, P. M. Zieliński, T. Wasiutyński, Y. Miyazaki, A. Inaba, D. Pinkowicz, B. Sieklucka, Magnetocaloric effect in $\{[Fe(pyrazole)_4]_2[Nb(CN)_8]\cdot 4H_2O\}_n$ molecular magnet, J. Magn. Magn. Mater. **354** (2013) 359-362, DOI: 10.1016/j.jmmm.2013.11.047

[109] M. Fitta, R. Pełka, M. Gajewski, M. Mihalik, M. Zentková, D. Pinkowicz, B. Sieklucka, M. Bałanda, Magnetocaloric effect and critical behavior in $Mn_2$-imidazole-$[Nb(CN)_8]$ magnetic sponge, J. Magn. Magn. Mater. **396** (2015) 1-8, DOI: 10.1016/j.jmmm.2015.07.112

[110] M. Gajewski, R. Pełka, M. Fitta, Y. Miyazaki, Y. Nakazawa, M. Bałanda, M. Reczyński, B. Nowicka, B. Sieklucka, Magnetocaloric effect of high-spin cluster with $Ni_9W_6$ core, J. Magn. Magn. Mater. **414** (2016) 25-31, DOI: 10.1016/j.jmmm.2016.04.062

[111] R. Pełka, M. Gajewski, Y. Miyazaki, S. Yamashita, Y. Nakazawa, M. Fitta, D. Pinkowicz, B. Sieklucka, Magnetocaloric effect in $Mn_2$-pyrazole-$[Nb(CN)_8]$ molecular magnet by relaxation calorimetry, J. Magn. Magn. Mater. **419** (2016) 435-441, DOI: 10.1016/j.jmmm.2016.06.074

[112] P. Konieczny, R. Pełka, D. Czernia, R. Podgajny, Rotating magnetocaloric effect in an anisotropic two-dimensional $Cu^{II}[W^V(CN)_8]^{3-}$ molecular magnet with topological phase



transition: Experiment and theory, Inorg. Chem. **56** (2017) 11971-11980, DOI: 10.1021/acs.inorgchem.7b01930

[113] P. Konieczny, Ł. Michalski, R. Podgajny, S. Chorąży, R. Pełka, D. Czernia, S. Buda, J. Mlynarski, B. Sieklucka, T. Wasiutyński, Self-enhancement of rotating magnetocaloric effect in anisotropic two-dimensional (2D) cyanido-bridged $Mn^{II}$-$Nb^{IV}$ molecular ferrimagnet, Inorg. Chem. **56** (2017) 2777-2783, DOI: 10.1021/acs.inorgchem.6b02941

[114] P. Konieczny, S. Chorąży, R. Pełka, K. Bednarek, T. Wasiutyński, S. Baran, B. Sieklucka, R. Podgajny, Double Magnetic relaxation and magnetocaloric effect in {$Mn_9$[W(CN)$_8$]$_6$(4,4'-dpds)$_4$} cluster-based network, Inorg. Chem. **56** (2017) 7089-7098, DOI: 10.1021/acs.inorgchem.7b00733

[115] M. Fitta, R. Pełka, W. Sas, D. Pinkowicz, B. Sieklucka, Dinuclear molecular magnets with unblocked magnetic connectivity: magnetocaloric effect, RSC Adv. **8** (2018) 14640-14645, DOI: 10.1039/c8ra01609g

[116] M. Fitta, R. Pełka, P. Konieczny, M. Bałanda, Multifunctional molecular magnets: magnetocaloric effect in octacyanometallates, Crystals **9** (2019) 9, DOI: 10.3390/cryst9010009

[117] R. Pełka, P. Konieczny, Y. Miyazaki, Y. Nakazawa, T. Wasiutyński, A. Budziak, D. Pinkowicz, B. Sieklucka, Comprehensive thermodynamic study of three Co(II)- and Fe(II)-based octacyanoniobates, Phys. Rev. B **104** (2021) 214428, DOI: 10.1103/PhysRevB.104.214428

[118] P. Konieczny, R. Pełka, E. Kuźniak, R. Podgajny, Anisotropy of magnetocaloric effect and rotating magnetocaloric effect in 2D coordination polymer based on {$Cu^{II}$[$W^V$(CN)$_8$]}$^-$ and adeninium cations, J. Magn. Magn. Mat. **563** (2022) 170001, DOI: 10.1016/j.jmmm.2022.170001

[119] R. Pełka, Y. Miyazaki, Y. Nakazawa, D. Pinkowicz, B. Sieklucka, Exploring magnetocaloric effect of coordination polymer based on Mn(II) and Nb(IV) by relaxation calorimetry, J. Phys. Chem. Solids **192** (2024) 112090, DOI: 10.1016/j.jpcs.2024.112090

[120] J. M. Herrera, P. Franz, R. Podgajny, M. Pilkington, M. Biner, S. Decurtins, H. Stoeckli-Evans, A. Neels, R. Garde, Y. Dromzée, M. Julve, B. Sieklucka, K. Hashimoto, S. Ohkoshi, M. Verdaguer, Three-dimensional bimetallic octacyanidometalates [$M^{IV}${(μ-CN)$_4$$Mn^{II}$(H$_2$O)$_2$}$_2$·4H$_2$O]$_n$ (M=Nb, Mo, W): Synthesis, single-crystal X-ray diffraction and magnetism, C. R. Chimie **11** (2008) 1192-1199, DOI: 10.1016/j.crci.2008.06.002

[121] J. A. Hofmann, A. Paskin, K. J. Tauer, R. J. Weiss, Analysis of ferromagnetic and antiferromagnetic second-order transitions, J. Phys. Chem. Solids **1** (1956) 45-60, DOI: 10.1016/0022-3697(56)90010-5

[122] J. Wucher, J. D. Wasscher, Couplage de spins et anomalies de chaleur spécifique aux temperatures de l'hélium et de l'hydrõgène liquids pour l'acétate complexe [Cr$_3$(CH$_3$COO)$_6$(OH)$_2$]Cl·8H$_2$O, Physica **20** (1954) 721-726, DOI: 10.1016/S0031-8914(54)80184-4

[123] D. A. Garanin, Extended Debye model for molecular magnets, Phys. Rev. B **78** (2008) 020405(R), DOI: 10.1103/PhysRevB.78.020405

[124] M. Sorai, A. Kosaki, H. Suga, S. Seki, Particle size effect on the magnetic and surface heat capacities of β-Co(OH)$_2$ and Ni(OH)$_2$ crystals between 1.5 and 300 K, J. Chem. Thermodyn. **1** (1969) 119-140, DOI: 10.1016/0021-9614(69)90052-4

[125] M. Sorai, S. Seki, Heat capacity of [Cr$_4$(OH)$_6$(en)$_6$](SO$_4$)$_3$·10H$_2$O from 1.4 to 200 K and spin interaction, J. Phys. Soc. Jpn. **32** (1972) 382-393, DOI: 10.1143/JPSJ.32.382

[126] J. Sólyom, Fundamentals of the Physics of Solids. Structure and Dynamics., Vol. 1, Springer Verlag, Berlin, 2007.

[127] S. Yamamoto, S. Brehmer, H.-J. Mikeska, Elementary excitations of Heisenberg ferrimagnetic spin chains, Phys. Rev. B **57** (1998) 13610, DOI: 10.1103/PhysRevB.57.13610

[128] N. Karchev, Towards the theory of ferrimagnetism, J. Phys.: Condens. Matter **20** (32) (2008) 325219, DOI: 10.1088/0953-8984/20/32/325219

[129] S. Ohkoshi, T. Iyoda, A. Fujishima, K. Hashimoto, Magnetic properties of mixed ferro-ferrimagnets composed of Prussian blue analogs, Phys. Rev. B **56** (1997) 11642, DOI: 10.1103/PhysRevB.56.11642

[130] H. Oesterreicher, F. T. Parker, Magnetic cooling near Curie temperatures above 300 K, J. Appl. Phys. **55** (1984) 4334-4338, DOI: 10.1063/1.333046

[131] V. Franco, A. Conde, J. M. Romero-Enrique, J. S. Blázquez, A universal curve for the magnetocaloric effect: an analysis based on scaling relations, J. Phys.: Condens. Matter **20** (2008) 285207, DOI: 10.1088/0953-8984/20/28/285207

[132] V. Franco, J. S. Blázquez, A. Conde, Field dependence of the magnetocaloric effect in materials with a second order phase transition: A master curve for the magnetic entropy change, Appl. Phys. Lett. **89** (2006) 222512, DOI: 10.1063/1.2399361

[133] M. Campostrini, M. Hasenbusch, A. Pelissetto, P. Rossi, E. Vicari, Critical exponents and equation of state of the three-diemnsional Heisenberg universality class, Phys. Rev. B **65** (2002) 144520, DOI: 10.1103/PhysRevB.65.144520

[134] A. Pelissetto, E. Vicari, Critical phenomena and renormalization-group theory, Phys. Rep. **368** (2002) 549-727, DOI: 10.1016/S0370-1573(02)00219-3

[135] A. Kitanovski, P. W. Egolf, Thermodynamics of magnetic refrigeration, Int. J. Refrig. **29** (2006) 3-21, DOI: 10.1016/j.ijrefrig.2005.04.007

[136] Y. Bingfeng, Z. Yan, G. Qiang, Y. Dexi, Research on performance of regenerative room temperature magnetic refrigeration cycle, Int. J. Refrig. **29** (2006) 1348-1357, DOI: 10.1016/j.ijrefrig.2006.07.015

[137] G. Diguet, G. Lin, J. Chen, Performance characteristics of magnetic Brayton refrigeration cycles using Gd, Gd$_{0.74}$Tb$_{0.26}$, and (Gd$_{3.5}$Tb$_{1.5}$)Si$_4$ as the working substance, Int. J. Refrig. **35** (2012) 1035-1042, DOI: 10.1016/j.ijrefrig.2011.12.004

[138] V. K. Pecharsky, K. A. Gschneidner Jr., A. O. Pecharsky, A. M. Tishin, Thermodynamics of the magnetocaloric effect, Phys. Rev. B **64** (2001) 144406, DOI: 10.1103/PhysRevB.64.144406